\documentclass[lettersize,journal]{IEEEtran}
\usepackage{amsmath,amsfonts,amssymb}
\usepackage{algorithm2e}
\usepackage{algpseudocode}
\RestyleAlgo{ruled}
\ifCLASSINFOpdf
 \usepackage[pdftex]{graphicx}
  \graphicspath{{./images/}}
\else
\fi
\usepackage{cite}

\makeatletter
\def\endthebibliography{%
  \def\@noitemerr{\@latex@warning{Empty `thebibliography' environment}}%
  \endlist
}
\makeatother

\usepackage[english]{babel}
\usepackage{amsthm}
\usepackage{hyperref}
\usepackage{cleveref}
\usepackage{lipsum}
\usepackage{caption}
\newtheorem{theorem}{Theorem}
\newtheorem{lemma}{Lemma}

\newtheorem{corollary}{Corollary}
\newtheorem*{remark}{Remark}
\theoremstyle{definition}

\allowdisplaybreaks[1]

\usepackage{color}
 \allowdisplaybreaks
\usepackage{soul}

\DeclareMathOperator*{\argmin}{argmin}
\DeclareMathOperator*{\argmax}{argmax}

\makeatletter
\newcommand*{\rom}[1]{\expandafter\@slowromancap\romannumeral #1@}
\makeatother

\begin{document}

\title{Analog-Digital Beam Focusing for Line of Sight Wide-aperture MIMO with  Spherical Wavefronts}

\author{\IEEEauthorblockN{Jiyoung~Yun, \IEEEmembership{Graduate Student Member, IEEE}, Hojun~Rho, \IEEEmembership{Graduate Student Member, IEEE}, and Wan Choi, \IEEEmembership{Fellow, IEEE}}
\thanks{J. Yun, H. Rho, and W. Choi are with the Department of Electrical and Computer Engineering and the Institute of New Media and Communications, Seoul National University (SNU), Seoul 08826, Korea (e-mail: \{jyyun423, ted1102, wanchoi\}@snu.ac.kr) (Corresponding author: Wan Choi.)}
}

\maketitle

\begin{abstract}
Enhancing high-speed wireless communication in the future relies significantly on harnessing high frequency bands effectively. These bands predominantly operate in line-of-sight (LoS) paths, necessitating well-configured antenna arrays and beamforming techniques for optimal spectrum utilization. Maximizing the potential of LoS multiple-input multiple-output (MIMO) systems, which are crucial for achieving high spectral efficiency, heavily depends on this. As the costs and power demands of mixed-signal devices in high frequency bands make a fully-digital architecture impractical for large-scale MIMO setups, our focus shifts to a hybrid analog-digital hardware configuration. Yet, analog processors' limitations restrict flexibility within arrays, necessitating a nuanced understanding of hardware constraints for optimal antenna configuration design. We explore array design that optimizes the spectral efficiency of hybrid systems, considering hardware constraints. We propose an optimal antenna configuration, leveraging the prolate matrix structure of the LoS channel between two planar arrays. Building on the optimal array configuration, we introduce a low-complexity explicit analog-digital beam focusing scheme that exploits the asymptotic behavior of the LoS channel matrix in the near-field region. Simulation results validate that the proposed antenna configuration and beam focusing scheme achieves near-optimal performance across a range of signal-to-noise ratios with low computational complexity, even under arbitrary rotations relative to the communication link.
\end{abstract}



\IEEEpeerreviewmaketitle

\section{Introduction}
Harnessing  high-frequency bands, including THz and sub-THz spectrum, is recognized as a pivotal technology for advanced communication systems beyond 5G. The expansive idle bandwidth within the high frequency bands holds significant potential for substantial increases in data rates \cite{akyildiz2014terahertz}. However, the high-frequency band faces challenges, primarily characterized by extremely high path loss which results in reduced communication range. To address this issue, both transmitters and receivers will be equipped with large-scale MIMO systems to compensate for the high path loss \cite{zhang2020prospective}. 

 A notable outcome of implementing large-scale MIMO and the subsequent reduction in communication range in the high-frequency band is the likelihood of communication occurring in the Fresnel region. Specifically, as the size of antenna arrays becomes larger compared to the wavelength and communication range, individual antennas gain improved signal resolvability without interference from multipath propagation, thanks to the beam focusing effect created by the spherical wavefronts of the signals. This phenomenon is commonly referred to as wide-aperture MIMO. Wide-aperture MIMO offers immediate benefits, particularly the potential for achieving high spatial multiplexing gains in MIMO systems, even under LoS conditions. This is especially relevant in high frequency band communications, where LoS transmission paths predominate as the wavelength decreases \cite{bohagen2009spherical, sun2014mimo, sarieddeen2019terahertz}.

Moreover, theoretical and empirical results support that careful design of transmitter (Tx) and receiver (Rx) arrays can optimize spatial multiplexing gain in near-field MIMO system. In particular, Gesbert et al. \cite{gesbert2002outdoor} introduced an optimal antenna spacing for uniform linear arrays (ULAs) in an LoS channel that maximizes spatial degree of freedom (DoF) while ensuring equal power parallel sub-channels. The optimal antenna spacing for LoS MIMO with ULA in the extremely low to high SNR region was explored in \cite{do2020reconfigurable}. In \cite{larsson2005lattice}, the optimal antenna spacing for LoS MIMO with parallel uniform planar arrays (UPAs) was investigated, and the authors of \cite{song2015spatial} proposed the optimal spacing for arbitrarily rotated UPAs for LoS MIMO. 

 Even with the optimal antenna spacing,  an appropriate beamforming scheme is pivotal for achieving high spatial multiplexing gain.  In this context, Torkildson et al. \cite{torkildson2011indoor} proposed a beamforming scheme using the array-of-subarrays transceiver architecture, while optimizing antenna spacing. Similarly, the authors of \cite{mojahedian2023precoding} presented a low-complexity precoder and combiner for a fully-digital transceiver architecture, exploiting the optimally spaced ULAs' channel structure for LoS MIMO. Recently, the authors of \cite{palaiologos2023non} proposed a non-uniform array configuration that is robust over a range of transmit distances in LoS MIMO. However, these approaches, while enhancing LoS MIMO channel capacity, are highly sub-optimal for systems with hardware constraints on Tx and Rx.

Achieving maximum channel capacity in wide-aperture MIMO systems with spherical wavefronts heavily hinges on signal processing hardware constraints, which significantly influence achievable spectral efficiency. In conventional MIMO systems with a limited number of antennas, digital signal processing is common in the baseband, with each array element linked to a dedicated radio frequency (RF) chain. However, the exorbitant cost and power consumption of RF chains render this approach impractical, especially in millimeter-wave (mmWave) and terahertz (THz) systems. To address this, high-frequency MIMO systems often adopt a hybrid analog-digital architecture that employs a phase-shifter between the antenna and the RF chain to compensate the lack of RF chains \cite{sohrabi2016hybrid}. More recently, dynamic metasurface antennas (DMAs) that enable programmable control of communication beam patterns were proposed \cite{shlezinger2021dynamic}. For such hybrid systems, signal processing begins with the analog processor, followed by reduced-dimensional processing in the baseband. Consequently, the achievable spectral efficiency of a hybrid MIMO system fundamentally differs from that of a fully-digital MIMO system. Designing hybrid beamforming strategies to maximize data rates poses a significant challenge due to the lack of a known closed-form solution for the hybrid precoder and combiner. Nevertheless, extensive research, including notable works such as \cite{ahmed2018survey, zhang2019hybrid}, has focused on hybrid beamforming to harness its benefits. However, prior studies have predominantly explored far-field beam design, emphasizing planar wavefronts rather than spherical ones.

For near-field beam focusing with a hybrid architecture, the authors of \cite{zhang2022beam} recently proposed a beam focusing-aware hybrid precoding scheme. This scheme, which considers spherical wavefronts in the near-field region, adopted iterative optimization algorithms to address a non-convex hybrid precoder design problem. However, while this work primarily addressed communication between a wide-aperture MIMO transmitter and single-antenna receivers in the Fresnel region, it did not fully exploit the distinctive channel characteristics of  high frequency band nor did it account for the presence of multiple antennas at the receiver. Specifically, the authors assumed a strong LoS channel equipped with a planar array but overlooked the potential of leveraging the doubly block Toeplitz structure of the corresponding channel. In \cite{wu2022distance}, a distance-aware phase-shifter-based hybrid precoding scheme was investigated. They proposed a transmitter capable of dynamically adjusting the number of active RF chains based on the degree of freedom (DoF) in the near-field region. While  they successfully enhanced data rates by exploiting additional DoFs in the near-field region, they fell short of fully capitalizing on the spatial multiplexing gain inherent in the underlying channel structure.

Motivated by the desire to enhance spatial multiplexing in hybrid LoS wide-aperture MIMO systems, our study delves into the maximum achievable spectral efficiency and the corresponding optimal antenna configuration, as well as transmit/receive beam patterns utilizing a hybrid antenna architecture. Our analysis sets forth an upper bound on the achievable rate of the general hybrid architecture, providing closed-form solutions for antenna spacing as dictated by array hardware constraints to achieve maximum spectral efficiency. Additionally, we propose a low-complexity design algorithm for a hybrid precoder and combiner in the near-field region, which nearly achieves the maximum spectral efficiency by leveraging the channel structure inherent in LoS wide-aperture MIMO systems. The key contributions of this paper are summarized as follows.

\noindent 1) Optimal configuration of antenna arrangement:
\begin{itemize}
  \item We figure out the optimal configuration of antenna arrangement tailored to the hardware constraints of  wide-aperture MIMO systems, focusing specifically on a two-dimensional antenna array within LoS-dominant high frequency systems employing a hybrid analog and digital signal processor.
  \item Our investigation focuses on determining the performance limits of the hybrid LoS MIMO system. We propose an optimal spacing between adjacent antennas and an optimal shape for the planar array, taking into account the unique characteristics of the high frequency spectrum such as THz.
\end{itemize}
2) Low-complexity analog-digital beam focusing scheme:
\begin{itemize}
\item We put forth a novel analog-digital beam focusing scheme designed to achieve high spectral efficiency in the optimally arranged wide-aperture MIMO system.
\item Leveraging the asymptotic characteristics of the LoS wide-aperture MIMO channel, we present a closed-form beam focusing solution. This solution significantly reduces the computational complexity associated with the joint construction of hybrid precoders and combiners.
\end{itemize}
3) Performance evaluation:
\begin{itemize}
\item We demonstrate that our proposed antenna arrangement scheme outperforms the existing schemes, providing a tangible performance gain while achieving near-optimal spectral efficiency.
\item Simulation results pertaining to the performance of the proposed beamforming strategy are presented. These results validate that our solution approaches its unconstrained performance limit, showcasing the effectiveness of the proposed approach in practical scenarios.
\end{itemize}

In summary, our contributions offer insights into optimizing the hardware and beamforming aspects of wide-aperture MIMO systems, with a focus on LoS-dominant high frequency scenarios. The proposed solutions are supported by thorough investigations and simulations, reinforcing their practical applicability and performance benefits.

This paper is structured as follows. In Section \ref{system_model}, we present the signal and channel model, offering the information-theoretic rate for both fully-digital and hybrid antenna architectures. Section \ref{optspacing} delves into the exploration of the maximum achievable rate and presents the array configuration that achieves the optimal rate. Following this, in Section \ref{beamforming}, we present an analog-digital beam focusing scheme. Our numerical results are detailed in Section \ref{results}. We draw conclusions in Section \ref{conclusion}.

Throughout the paper, lower- and upper-case bold letters represent vector and matrix, respectively. The notation $\mathbf{x} \sim \mathcal{CN}(\boldsymbol{\mu}, \mathbf{C})$ means that the random vector $\mathbf{x}$ follows a complex Gaussian distribution with mean $\boldsymbol{\mu}$ and covariance matrix $\mathbf{C}$. $[\mathbf{A}]_{i,k}$ denotes $(i,k)$-th element of the matrix $\mathbf{A}$, $\lVert \mathbf{A} \rVert$ denotes the determinant of the matrix $\mathbf{A}$ and $\mathbf{A}^{*}$ is conjugate transpose of the matrix $\mathbf{A}$. We denote $[\mathbf{A} \parallel \mathbf{b}]$ to append a vector $\mathbf{b}$ to matrix $\mathbf{A}$. We use $\left[a\right]^{+}$ to denote the maximum between $a$ and $0$, i.e., $\left[a\right]^{+} = \max (a,0)$. $\lfloor\cdot\rfloor$ is a floor function. Set is denoted by $\{\cdot\}$ and $\otimes$ denotes the Kronecker product. $\#(\cdot)$ denotes the number of integer included in the interval. 

\section{System Model} \label{system_model}
In this section, we delineate the signal and channel model for LoS MIMO communication employing spherical wavefronts. Subsequently, we articulate an optimization problem aimed at designing the antenna configuration, transmission, and reception beam patterns to maximize data rate in near-field communications. Our focus lies on a single-user LoS MIMO system featuring corresponding UPAs comprising $M$ transmit and $N$ receive antennas, all of which can be arbitrarily oriented in space.

\begin{figure*}[t!]
    \centering
\captionsetup{justification=centering}
    \includegraphics[scale=0.6]{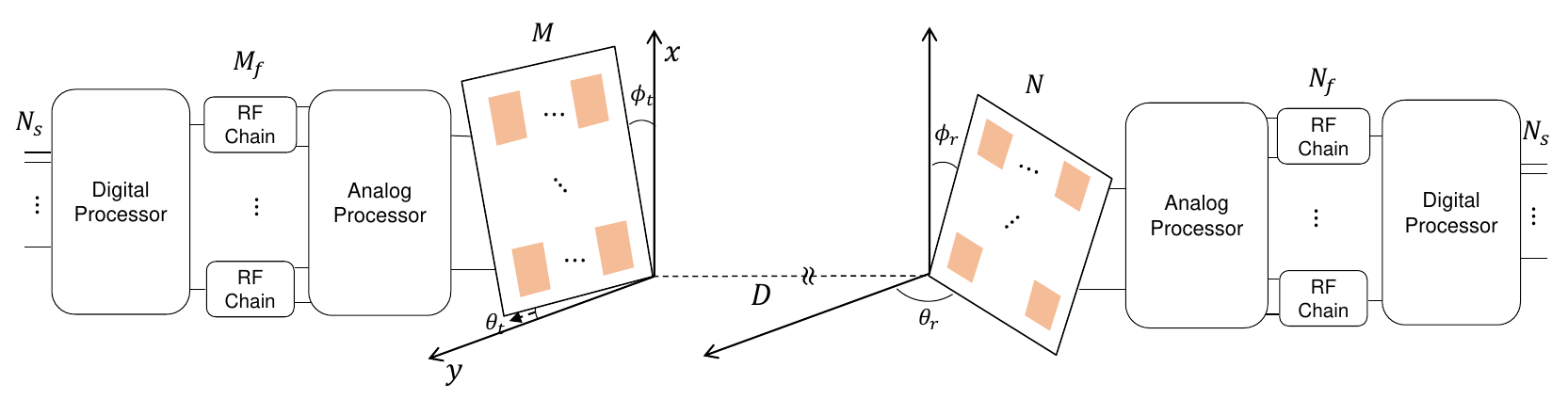}
    \caption{ Generic UPA wide-aperture MIMO system with hybrid architecture.} 
    \label{fig:systemmodel}
\end{figure*}

\subsection{Near-Field Channel Model}
We assume that both the receiver and the transmitter deployed a planar antenna array with $N_{\text{v}}$ and $M_{\text{v}}$ elements in the vertical direction, and $N_{\text{h}}$ and $M_{\text{h}}$ elements in the horizontal direction, i.e., $N=N_{\text{v}}\times N_{\text{h}}$, and $ M=M_{\text{v}}\times M_{\text{h}}$. We denote the uniform antenna spacing across the vertical and horizontal direction between adjacent Tx antennas as $d_{\text{t,v}}$, and $d_{\text{t,h}}$, respectively, and between Rx antennas as $d_{\text{r,v}}$, and $d_{\text{r,h}}$. As depicted in Fig. \ref{fig:systemmodel}, the geometric relationship between two opposing planar arrays can be specified by four 3D rotation angles: $\theta_{\text{t}}$, $\phi_{\text{t}}$, $\theta_{\text{r}}$, and $\phi_{\text{r}}$. The geometric location of the $m$-th Tx antenna is denoted as $\textbf{t}_{m}$ and the $n$-th Rx antenna as $\textbf{r}_{n}$, both are represented by the $\text{(x,y,z)}$ Cartesian coordinates, with the $z$-axis representing the direction of the communication link. We define a matrix $\mathbf{T} \in \mathbb{R}^{3 \times M}$, wherein each column represents the Cartesian coordinates of the $m$-th Tx antenna $\textbf{t}_{m}$. The antenna location matrix $\mathbf{R} \in \mathbb{R}^{3 \times N}$ is defined similarly to $\mathbf{T}$. The Tx and Rx array apertures are denoted as $L_{\text{t}}$ and $L_{\text{r}}$, respectively, defined as the largest Euclidean distance between two elements of the array \cite{palaiologos2023non}.
\begin{align} \label{eqn:aperture}
    L_{\text{t}} = \max_{i,j} \lVert \textbf{t}_{i} - \textbf{t}_{j} \rVert_{2}, \forall i,j \in \{0,\ldots,M-1\}
\end{align}
While we omit the definition of $L_{\text{r}}$, it is similarly defined to $L_{\text{t}}$. The channel between the $m$-th transmit and $n$-th receive antenna is characterized by
\begin{align}
h_{n,m}=\frac{\sqrt{G_tG_r} \lambda_{c} }{4\pi d_{n,m}}e^{-j \frac{2\pi}{\lambda_{c}} d_{n,m}} \qquad &n=0,\cdots,N-1 \\
&m=0,\cdots,M-1 \nonumber
\end{align}
where $d_{n,m}$ is the distance between the $m$-th transmit antenna and the $n$-th receive antenna, i.e. $d_{n,m} = \lVert \textbf{r}_{n} - \textbf{t}_{m} \rVert_{2}$, $\lambda_{c}$ is the carrier wavelength, and $G_t$ and $G_r$ represent the antenna gains. We consider the approximately constant path loss across different pairs of antennas \cite{do2020reconfigurable}, which holds true for the practical LoS MIMO system where the array aperture is smaller than the communication distance $D$, i.e., $L_{\text{t}},L_{\text{r}} \ll D$. Consequently, our attention shifts towards the phase shifts within the channel matrix, as they determine the spatial multiplexing capabilities of a LoS MIMO system. The corresponding normalized channel matrix, denoted as $\mathbf{H}$, is expressed as
\begin{align} \label{eqn:H}
    \mathbf{H} = \begin{bmatrix}
        e^{-j \frac{2\pi}{\lambda_{c}} d_{0,0}} & \cdots & e^{-j \frac{2\pi}{\lambda_{c}} d_{0,M-1}} \\
        \vdots & \ddots & \vdots \\
        e^{-j \frac{2\pi}{\lambda_{c}} d_{N-1,0}} & \cdots & e^{-j \frac{2\pi}{\lambda_{c}} d_{N-1,M-1}}
    \end{bmatrix},
\end{align} 
where $\lVert \mathbf{H} \rVert^{2}_{F}=NM$. Let the discrete-time transmitted signal be $\mathbf{x}=\mathbf{F}\mathbf{s}$, where $\mathbf{s}$ is a vector consisting of unit-power symbols, $\mathbf{F}$ is a transmit beamforming matrix, and the transmit signal's power is constrained to be less than or equal to $P_{\text{t}}$. The received signal $\mathbf{y}$ is given as
\begin{align} \label{eqn:y}
    \mathbf{y} = \sqrt{\zeta}\mathbf{H}\mathbf{x} + \mathbf{z},
\end{align}
where $\zeta$ denotes the approximated path loss, i.e., $\sqrt{\zeta} = \frac{\sqrt{G_tG_r}\lambda_{c}}{4\pi D}$, and $\mathbf{z}$ is an additive white Gaussian noise with zero mean and unit variance denoted as $\sigma_{n}^{2} \mathbf{I}$.

The antenna distance $d_{n,m}$ between the $m$-th transmit antenna and the $n$-th receive antenna is calculated as
\begin{align} \label{eqn:exact_r}
    d_{n,m}=\sqrt{(D+\text{r}^{z}_{n} - \text{t}^{z}_{m})^2+(\text{r}^{x}_{n}-\text{t}^{x}_{m})^2+(\text{r}^{y}_{n}-\text{t}^{y}_{m})^2},
\end{align}
with the slight abuse of notation $D+\text{r}^{z}_{n}$ correspond to the $z$ coordinate of the $n$-th receive antenna. Given the assumption that the communication distance is significantly greater than the array aperture, we may apply a Taylor series expansion in \eqref{eqn:exact_r} and express as 
\begin{align}
    d_{n,m} \approx D + \text{r}^{z}_{n} - \text{t}^{z}_{m} + \frac{(\text{r}^{x}_{n}-\text{t}^{x}_{m})^2}{2D} + \frac{(\text{r}^{y}_{n}-\text{t}^{y}_{m})^2}{2D}.
\end{align}

The channel matrix in \eqref{eqn:y} with the approximated antenna distance is equivalent as
\begin{align} \label{eqn:approxH}
    \mathbf{H} \approx  \mathbf{D}^{*}_{\text{r}} \tilde{\mathbf{H}} \mathbf{D}_{\text{t}},
\end{align}
where $\mathbf{D}_{\text{r}}$ and $\mathbf{D}_{\text{t}}$ are diagonal matrices representing the phase shifts caused by the receiver and transmitter independently. Specifically, 
\begin{align*}
[\mathbf{D}_{\text{t}}]_{m,m}&=e^{j \frac{2\pi}{\lambda_{c}} (\text{t}^{z}_{m} - \frac{(\text{t}^{x}_{m})^2 + (\text{t}^{z}_{m})^2}{2D})},\\
[\mathbf{D}_{\text{r}}]_{n,n}&= e^{j \frac{2\pi}{\lambda_{c}} (D+\text{r}^{z}_{n}+ \frac{(\text{r}^{x}_{n})^2 + (\text{r}^{y}_{n})^2}{2D})},
\end{align*} and $\tilde{\mathbf{H}}$ is the channel matrix contributed by spatial multiplexing on the orthogonal plane relative to the link direction, given by $$[\tilde{\mathbf{H}}]_{n,m} = e^{j \frac{2 \pi}{\lambda_{c}} \frac{\text{r}^{x}_{n} \text{t}^{x}_{m} + \text{r}^{y}_{n} \text{t}^{y}_{m}}{D}}.$$

\subsection{Signal Model}
The successful transmission and reception of the signal in \eqref{eqn:y} hinge upon the signal processing capabilities embedded in the antenna architecture. Our focus lies in achieving focused beams within a linear MIMO system while operating within the constraints of a limited number of RF chains. Although a fully digital MIMO system, where each antenna possesses a dedicated RF chain, offers highly flexible signal processing capabilities by enabling individual processing of inputs to each element, the feasibility of such a setup is hindered by the cost and power consumption associated with high frequency signals such as THz. Instead, a hybrid architecture, which combines reduced-dimensional digital signal processing with analog processing, emerges as a more practical approach for wide-aperture MIMO systems. This hybrid architecture integrates digital signal processing with a controlled level of analog signal processing. The prevailing hybrid beamforming architecture, extensively examined in existing literature, involves incorporating a phase-shifter layer between the RF chain and the antennas. More recently, the DMA architecture, leveraging radiating metamaterial elements, has been proposed in \cite{shlezinger2021dynamic}.  In our system model for hybrid architecture, the transmitter is equipped with $M_{\text{f}}$ RF chains to transmit $N_s$ data streams, where $N_s$ satisfies $N_s \le M_{\text{f}} \le M$.  Similarly, the receiver is equipped with $N_{\text{f}}$ RF chains to receive $N_s$ data streams. In a system constrained by the number of RF chains, the transmitter and receiver can communicate up to $\min(N_{\text{f}}, M_{\text{f}})$ data streams, thus imposing a limit on the overall achievable rate.

\subsection{Problem Formulation}
Throughout the paper, we assume perfect knowledge of the channel $\mathbf{H}$ by both the transmitter and receiver. In fact, estimating the near-field channel accurately with a limited number of RF chains is challenging and has been explored in the literature \cite{cui2021near, dovelos2021channel, zhang2023near, zhang2023near2}. We make the assumption of perfect channel knowledge for analytical purposes, and considering imperfect channel knowledge is beyond the scope of this study. For a hybrid antenna array that transmits $N_s$ data streams, the spectral efficiency is given by \cite{goldsmith2003capacity}
\begin{align} \begin{split} \label{eqn:hybridR}
    R = \log \det \Big(  \mathbf{I}_{N_s} + \frac{\zeta}{N_{s} \sigma_{n}^{2}} \mathbf{R}^{-1} \mathbf{W}^{*} \mathbf{H} \mathbf{F}\mathbf{F}^{*} \mathbf{H}^{*} \mathbf{W}\Big),
\end{split}
\end{align}
where $\mathbf{W}$ is a receiver combining matrix and $\mathbf{R}=\mathbf{W}^{*}\mathbf{W}$ is a noise covariance matrix after combining. For the RF-chain limited MIMO architecture, the maximum spectral efficiency can be achieved by jointly optimizing the precoder, the combiner, and the antenna configuration. Note that for the strong LoS channel, the channel matrix $\mathbf{H}$ only depends on the transmitter and the receiver array configurations.  Thus, we focus on the optimization problem 
\begin{subequations}
\begin{align}
    \mathcal{P}_1 :&\max_{\mathbf{H}, \mathbf{F},  \mathbf{W}} R \\
    &~~\textrm{s.t.} \, \,  \lVert \mathbf{H} \rVert _{\text{F}}^{2} = NM, \\
    &~~~~~~\text{Tr} (\mathbf{F} \mathbf{F}^{*}) \le P_{\text{t}}, \\
    &~~~~~~\mathbf{W} \in \mathcal{W}, \mathbf{F} \in \mathcal{F}, \\
    &~~~~~~L_{\text{t}} = \max_{i,j} \lVert \textbf{t}_{i} - \textbf{t}_{j} \rVert_{2}, \forall i,j \in \{0,\ldots,M-1\} ,\\
    &~~~~~~L_{\text{r}} = \max_{i,j} \lVert \textbf{r}_{i} - \textbf{r}_{j} \rVert_{2}, \forall i,j \in \{0,\ldots,N-1\},
\end{align} \label{eqn:maxR}
\end{subequations}
where $\mathcal{W}$ and $\mathcal{F}$ denotes the feasible combiner and precoder set, respectively, which depend on the underlying hardware architecture.
\section{Optimal Spatial Multiplexing Gain by Analog-Digital Beam Focusing with  Generic Planar Arrays} \label{optspacing}

In this section, we explore optimal spatial multiplexing for planar arrays equipped with a hybrid architecture. Configuring the arrangement of the array elements for a strong LoS channel is crucial for maximizing spectral efficiency. Channel matrices lacking specific geometrical arrangements are likely to be rank-deficient, resulting in undesirable low data rates  even over short communication distances of just a few meters \cite{wu2022distance}. Thus, our focus is on the antenna array arrangement that maximizes the spectral efficiency of the hybrid MIMO channel.

Configuring the maximum achievable rate of the hybrid architecture is generally impossible since the constraints of the precoder and the combiner, i.e. $\mathcal{F}$ and $\mathcal{W}$, are non-convex, making the maximization problem of \eqref{eqn:maxR} non-convex. However, by relaxing the non-convex constraint of the analog precoder and combiner, the upper bound of \eqref{eqn:maxR} can be achieved. Consider that the number of transmit data streams is smaller than or equal to the rank of the channel, i.e., $N_s \le \text{rank}(\mathbf{H})$, while the transmitter and receiver are equipped with fully digital signal processors. Maximum spectral efficiency can be achieved by setting the precoder and the combiner according to the singular value decomposition (SVD), $[\mathbf{V}]_{1:N_s}$ and $[\mathbf{W}]_{1:N_s}$, $\mathbf{U}^{*}$ where the SVD of the channel matrix is given as $\mathbf{H} = \mathbf{U} \mathbf{\Sigma} \mathbf{V}^{*}$. Thus, we aim at maximizing the upper bound of the spectral efficiency in \eqref{eqn:hybridR} by considering the following equivalent problem:
\begin{subequations}
\begin{align}
    \mathcal{P}_2 :\max_{\boldsymbol{\lambda}, \mathbf{p}} &\sum_{i=1}^{N_s} \log (1+ \frac{\zeta}{\sigma_{n}^{2}} \lambda_i p_i)\\
    \textrm{s.t.} ~~ &\lVert \boldsymbol{\lambda} \rVert_{1} = NM,\\
    &\lVert \mathbf{p} \rVert_{1} = P_{\text{t}},
\end{align} \label{eqn:P2}
\end{subequations} 
where $\boldsymbol{\lambda} = \left(\lambda_1, \lambda_2, \ldots, \lambda_{\min(N,M)}\right)$ denotes the eigenvalues of the matrix $\mathbf{H}^{*}\mathbf{H}$ and the $i$-th element of the vector $\mathbf{p} \in \mathbb{C}^{M\times 1}$ denotes the power allocation for $i$-th symbol. It is worth noting that $\boldsymbol{\lambda}$ is a function of the Tx and Rx array in case of LoS system, thus the achievable rate can be significantly enhanced by appropriately configuring the array aperture. With a slight abuse of notation, we represent $\boldsymbol{\lambda}$ as the eigenvalues of the channel matrix $\mathbf{H}$ throughout this paper. Furthermore, it is well known that the optimal power allocation that maximizes the \eqref{eqn:P2} can be derived by the water-filling algorithm. That is, the upper bound of the spectral efficiency in the hybrid MIMO system \eqref{eqn:hybridR} is a function of the $N_s$ leading eigenvalues of the channel matrix $\mathbf{H}$.

The upper bound of the spectral efficiency in the hybrid MIMO system, which is formulated in $\mathcal{P}_2$, is maximized when the channel gain $\lVert \mathbf{H} \rVert^{2}_{F}$ is spread over $N_s$ parallel sub-channels while the sub-channel gains of the remaining $\min(N,M)-N_s$ sub-channels are zero. Put differently, the channel gain should be concentrated on the available sub-channels. Additionally, it is known that to maximize the spatial multiplexing gain in \eqref{eqn:P2} in the moderate to high SNR regime, every non-zero eigenvalues should be of equal magnitude. Consequently, the upper bound of spectral efficiency is given as
\begin{align} \label{eqn:maxmaxSE}
    \max_{\boldsymbol{\lambda}, \mathbf{p}} &\sum_{i=1}^{N_s} \log (1+\lambda_i p_i)= N_s \log (1+ \frac{\zeta P_{\text{t}} }{\sigma_{n}^{2}} \cdot \frac{N M}{N^{2}_s}).
\end{align}
It is important to highlight that the converse in \eqref{eqn:maxmaxSE} serves as an upper limit on data rates for all antenna systems equipped with a limited number of RF chains, including phase-shifter-based hybrid transceivers and DMA-based hybrid transceivers.

\begin{figure}[t!]
    \centering
    \includegraphics[scale=0.5]{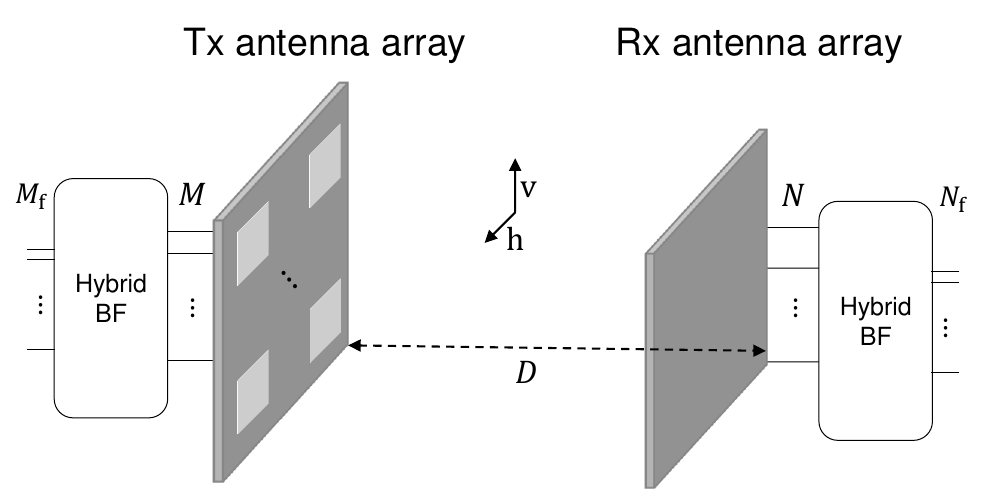}
    \caption{Parallel UPA wide-aperture MIMO system with hybrid architecture.}
    \label{fig:parasystemmodel}
\end{figure} 

In order to find the optimal antenna spacing that achieves \eqref{eqn:maxmaxSE}, we exploit the fact that the eigenvalues of $\mathbf{H}$ and $\tilde{\mathbf{H}}$ coincide, 
\begin{align} \begin{split} \label{eqn:SVD}
    \mathbf{H} &\approx  \textcolor{black}{\mathbf{D}^{*}_{r}} \tilde{\mathbf{H}} \mathbf{D}_{t} \\ &= \textcolor{black}{\mathbf{D}^{*}_{r}} \tilde{\mathbf{U}} \tilde{\mathbf{\Sigma}} \tilde{\mathbf{V}}^{*} \mathbf{D}_{t} = \mathbf{U} \mathbf{\Sigma}\mathbf{V}^{*}.
\end{split}
\end{align}
We initially focus on the parallel UPA where the rotation angles are zero, i.e. $\theta_t=\phi_t=\theta_r=\phi_r=0$. The matrix $\tilde{\mathbf{H}}$ with parallel UPAs can be expressed as the Kronecker product of two linear array channel matrices \cite{larsson2005lattice},
\begin{align} \label{eqn:paraH}
    \tilde{\mathbf{H}}^{\parallel} = \mathbf{H}_{\text{linv}} \otimes \mathbf{H}_{\text{linh}},
\end{align}
where $\mathbf{H}_{\text{linv}} \in \mathbb{C}^{N_{\text{v}} \times M_{\text{v}}}$ and $\mathbf{H}_{\text{linh}} \in \mathbb{C}^{N_{\text{h}} \times M_{\text{h}}}$ representing the channel between linear transmitter and receiver.  The two-dimensional array channel matrix can be decomposed into two MIMO sub-matrices, as illustrated in Fig. \ref{fig:parasystemmodel}. Let $\mathbf{G}_{\text{t}}=\tilde{\mathbf{H}}^{*} \tilde{\mathbf{H}}$ denote the channel gain matrix, which the eigenvalues of $\mathbf{G}_{\text{t}}$ being equal to that of $\tilde{\mathbf{H}}$. The channel gain matrix of the parallel UPA is given as
\begin{align} \label{eqn:GG}
    \mathbf{G}^{\parallel}_{\text{t}} &= (\mathbf{H}_{\text{linv}} \otimes \mathbf{H}_{\text{linh}})^{*} (\mathbf{H}_{\text{linv}} \otimes \mathbf{H}_{\text{linh}}) \\
    &=(\mathbf{H}^{*}_{\text{linv}}\mathbf{H}_{\text{linv}}) \otimes (\mathbf{H}^{*}_{\text{linh}}\mathbf{H}_{\text{linh}}) \\
    &= \mathbf{G}_{\text{linv}} \otimes \mathbf{G}_{\text{linh}},
\end{align}
where $\mathbf{G}_{\text{linv}} \triangleq (\mathbf{H}^{*}_{\text{linv}}\mathbf{H}_{\text{linv}})$ and $\mathbf{G}_{\text{linh}} \triangleq (\mathbf{H}^{*}_{\text{linh}}\mathbf{H}_{\text{linh}})$ are channel gain matrices corresponding to $\mathbf{H}_{\text{linv}}$ and $\mathbf{H}_{\text{linh}}$, respectively. The $(i,k)$ element of the channel gain matrices $\mathbf{G}_{\text{linv}}$ is given by
\begin{align} \label{eqn:GlinV}
    [\mathbf{G}_{\text{linv}}]_{i,k} = e^{-j\pi\Delta_{\text{v}}\frac{i-k}{N_{\text{v, max}}}(N_{r,v}-1)} \frac{\sin\left(\pi \Delta_{\text{v}} \frac{i-k}{N_{\text{v, max}}} N_{\text{v}}\right)}{\sin\left(\pi \Delta_{\text{v}} \frac{i-k}{N_{\text{v, max}}}\right)},
\end{align}
where $N_{\text{v, max}}$ denotes the maximum value between $N_{\text{v}}$ and $M_{\text{v}}$. We have defined the parameter $\Delta_{\text{v}} = \frac{d_{r,v}d_{t,v}N_{\text{v, max}}}{\lambda_{c} D}$ that depends on the antenna configuration to simplify the notation. The channel gain matrix $\mathbf{G}_{\text{linh}}$ is given similarly as $\mathbf{G}_{\text{linv}}$, with $N_{\text{h, max}}$ and $\Delta_{\text{h}}$ defined similarly to $N_{\text{v, max}}$ and $\Delta_{\text{v}}$, respectively. As the structures of two sub-matrices, $\mathbf{H}_{\text{linv}}$ and $\mathbf{H}_{\text{linh}}$, are identical, we omit the detailed derivation of $\mathbf{G}_{\text{linh}}$. As illustrated in \eqref{eqn:GlinV}, the channel gain matrix $\mathbf{G}_{\text{linv}}$ is a product of an unitary phase factor and a scaled prolate matrix, where the prolate matrix $\mathbf{B}(\alpha,K) \in \mathbb{C}^{\alpha \times \alpha}$ has entries defined as
\begin{align}
    [\mathbf{B}(\alpha,K)]_{i,k} = \frac{1}{\alpha} \frac{\sin \left(\pi (i-k) (K+1)/\alpha \right)}{\sin \left(\pi (i-k)/\alpha \right)}.
\end{align}
Thus, the element $(i,k)$ of $\mathbf{G}_{\text{linv}}$ can be expressed in terms of the prolate matrix as
\begin{align}
\begin{split}
    \textcolor{black}{\left[\mathbf{G}_{\text{linv}} \right]_{i,k}} = &e^{-j\pi\Delta_{\text{v}}\frac{(i-k)(N_{\text{v}}-1)}{N_{\text{v, max}}}} \\
    &\cdot \frac{N_{\text{v, max}}}{\Delta_{\text{v}}} \left[\mathbf{B}_{M_{\text{v}}}\left(\frac{N_{\text{v, max}}}{\Delta_{\text{v}}},N_{\text{v}} -1\right)\right]_{i,k},
\end{split}
\end{align} 
where $\mathbf{B}_{\beta}\left(\alpha,K\right)$ denotes the truncated prolate matrix $\mathbf{B}_{\beta}(\alpha,K)=[\mathbf{B}(\alpha,K)]_{1:\beta,1:\beta}$. As the eigenvalue distribution is invariant under unitary phase factors, the eigenvalue problem of $\mathbf{G}_{\text{linv}}$ boils down to the eigenvalue problem of the corresponding prolate matrices. Fortunately, the eigenvalue distribution of the truncated prolate matrix has recently been studied in \cite{zhu2017eigenvalue}.  In particular, \cite[Theorem 1]{zhu2017eigenvalue} establishes the spectrum concentration property for $\mathbf{B}_{\beta}(\alpha,K)$, indicating that the first $2\left\lfloor\frac{(2K+1)\beta}{2\alpha}\right\rfloor$ eigenvalues form a cluster around one, while the remaining eigenvalues tend to zero. Let $\omega_{0,(i)},~ \ldots, ~\omega_{\beta-1,(i)}, i \in \{\text{v,h}\}$ be the ordered eigenvalues of the normalized channel gain matrices $\mathbf{G}_{\text{lin}(i)}, i \in \{\text{v},\text{h}\}$. Then, the result of \cite{zhu2017eigenvalue}  is translated into the following lemma. 
\begin{lemma} \label{lma:lma1}
    For $i \in \{\mathrm{v},\mathrm{h}\}$, the normalized channel gain matrix $\frac{\Delta_i}{N_{i,\mathrm{max}}} \mathbf{G}_{\mathrm{lin}(i)}$, for any $M_{i}, N_{i}, \Delta_i$ such that $\frac{N_{i,\mathrm{max}}}{\Delta_i}\in\mathbb{N}$, suppose $M_{i}, N_{i} \le \frac{N_{i,\mathrm{max}}}{\Delta_i}$. Then, for any $\epsilon \in (0,\frac{1}{2})$,
    \begin{align}
    \begin{split}
        \omega_{2\left\lfloor\frac{\Delta_i}{2} N_{i,\mathrm{min}}\right\rfloor-R(M_{i},N_{i},\Delta_i,\epsilon),(i)} \ge 1-\epsilon, \\ \omega_{2\left\lfloor\frac{\Delta_i}{2} N_{i,\mathrm{min}}\right\rfloor+R(M_{i},N_{i},\Delta_i,\epsilon)+1,(i)} \le \epsilon,
    \end{split}
    \end{align} and
    \begin{align}
        \#(l: \epsilon \le \omega_{l,(i)} \le 1-\epsilon) \le 2R(M_{i},N_{i},\Delta_i,\epsilon)
    \end{align}
    where
    \begin{align*}
        R(N_{i},M_{i},\Delta_i,\epsilon) = \left( \frac{4}{\pi^2} \log (8M_{i}) + 6 \right) \log \left(\frac{16}{\epsilon} \right) \\ + 2\left[ \frac{-\log\left(\frac{\pi}{32} \epsilon \left( \left(\frac{N_{i,\mathrm{max}}}{M_{i}\Delta_{i}}\right)^2-1 \right) \right)}{\log\left(\frac{N_{i,\mathrm{max}}}{M_{i}\Delta_{i}}\right)}\right]^{+}
    \end{align*}
and $N_{\mathrm{v, min}} \triangleq \min(N_{\mathrm{v}},M_{\mathrm{v}})$, $N_{\mathrm{h, min}} \triangleq \min(N_{\mathrm{h}},M_{\mathrm{h}})$.
\end{lemma}

It should be noted that $R(N_{i},M_{i},\Delta_i,\epsilon)$ is a variable related to the number of eigenvalues that are neither clustered near $1$ nor clustered near $0$, which we represent as the \emph{transition band}. The size of the transition band scales as $\mathcal{O}\left(\log\frac{1}{\epsilon} \log N_{i,\text{min}}\right)$, which is negligible for the wide range of $\epsilon$ and $ N_{i,\text{min}}$  \cite{zhu2017eigenvalue}.

Directly following from Lemma \ref{lma:lma1}, the rank of the channel gain matrices can be approximated as $\text{rank}(\mathbf{G}_{\text{linv}})  = 2\left\lfloor\frac{\Delta_{\text{v}}}{2} N_{\text{v, min}}\right\rfloor $ and $\text{rank}(\mathbf{G}_{\text{linh}})  = 2\left\lfloor\frac{\Delta_{\text{h}}}{2} N_{\text{h},\text{min}}\right\rfloor $ and the non-zero eigenvalues of each matrices are clustered around $M_{\text{v}} N_{\text{v}} / 2\left\lfloor\frac{\Delta_{\text{v}}}{2} N_{\text{v, min}}\right\rfloor$ and $M_{\text{h}} N_{\text{h}} / 2\left\lfloor\frac{\Delta_{\text{h}}}{2} N_{\text{h, min}}\right\rfloor$, respectively. Taking advantage of the fact that the channel gain matrix of the parallel UPA $\mathbf{G}^{\parallel}_{\text{t}}$ is equivalent to the Kronecker product of $\mathbf{G}_{\text{linv}}$ and $\mathbf{G}_{\text{linh}}$, we can obtain the eigenvalue distribution of $\mathbf{G}^{\parallel}_{\text{t}}$ with the eigenvalue distribution of the matrices $\mathbf{G}_{\text{linv}}$ and $\mathbf{G}_{\text{linv}}$. Specifically, the rank of the matrix $\mathbf{G}^{\parallel}_{\text{t}}$ is the same as the number of non-zero eigenvalues, and the number of non-zero eigenvalues of $\mathbf{G}^{\parallel}_{\text{t}}$ is the same as the product of the number of non-zero eigenvalues of $\mathbf{G}_{\text{linv}}$ and $\mathbf{G}_{\text{linh}}$. 

\begin{corollary} \label{thm:cor1}
The number of non-zero eigenvalues of $\mathbf{G}^{\parallel}_{t}$ is approximated as $4\left\lfloor\frac{\Delta_{\mathrm{v}}}{2} N_{\mathrm{v},\mathrm{min}}\right\rfloor \left\lfloor\frac{\Delta_{\mathrm{h}}}{2} N_{\mathrm{h},\mathrm{min}}\right\rfloor $, where each non-zero eigenvalue forms a cluster around the value $\frac{MN}{4\left\lfloor\frac{\Delta_{\mathrm{v}}}{2} N_{\mathrm{v},\mathrm{min}}\right\rfloor \left\lfloor\frac{\Delta_{\mathrm{h}}}{2} N_{\mathrm{h},\mathrm{min}}\right\rfloor}$.
\end{corollary} 
\begin{IEEEproof}
    Let $\mathbf{Q}_{i}, i \in \{\mathrm{v, h}\}$ be the matrix composed of the eigenvectors of $\mathbf{G}_{\text{lin}(i)}$. Capitalizing on the relationship in \eqref{eqn:GG} we can derive, 
    \begin{align}
    \mathbf{G}^{\parallel}_{\text{t}} &= \mathbf{G}_{\text{linv}} \otimes \mathbf{G}_{\text{linh}}\\
    \begin{split}
    &\approx \mathbf{Q}_{\text{v}} \begin{pmatrix} \frac{M_{\text{v}} N_{\text{v}} }{2\left\lfloor\frac{\Delta_{\text{v}}}{2} N_{\text{v, min}}\right\rfloor}\mathbf{I}_{2\left\lfloor\frac{\Delta_{\text{v}}}{2} N_{\text{v, min}}\right\rfloor} & \mathbf{0} \\ \mathbf{0} & \mathbf{0} \end{pmatrix} \mathbf{Q}^{*}_{\text{v}} \\
    &\quad \otimes \mathbf{Q}_{\text{h}} \begin{pmatrix} \frac{M_{\text{h}} N_{\text{h}} }{2\left\lfloor\frac{\Delta_{\text{h}}}{2} N_{\text{h, min}}\right\rfloor}\mathbf{I}_{2\left\lfloor\frac{\Delta_{\text{h}}}{2} N_{\text{h, min}}\right\rfloor} & \mathbf{0} \\ \mathbf{0} & \mathbf{0} \end{pmatrix} \mathbf{Q}^{*}_{\text{h}}
    \end{split}\\
    \begin{split}
    &= \left(\mathbf{Q}_{\text{v}} \otimes \mathbf{Q}_{\text{h}}\right) \mathbf{\Lambda} \left(\mathbf{Q}^{*}_{\text{v}} \otimes \mathbf{Q}^{*}_{\text{h}}\right) 
    \end{split},
    \end{align}
    where
    \begin{align*}
        \mathbf{\Lambda} &= \begin{pmatrix} \frac{M_{\text{v}} N_{\text{v}} }{2\left\lfloor\frac{\Delta_{\text{v}}}{2} N_{\text{v, min}}\right\rfloor}\mathbf{I}_{2\left\lfloor\frac{\Delta_{\text{v}}}{2} N_{\text{v, min}}\right\rfloor} & \mathbf{0} \\ \mathbf{0} & \mathbf{0} \end{pmatrix} \\
        &\quad \otimes \begin{pmatrix} \frac{M_{\text{h}} N_{\text{h}} }{2\left\lfloor\frac{\Delta_{\text{h}}}{2} N_{\text{h, min}}\right\rfloor}\mathbf{I}_{2\left\lfloor\frac{\Delta_{\text{h}}}{2} N_{\text{h, min}}\right\rfloor} & \mathbf{0} \\ \mathbf{0} & \mathbf{0} \end{pmatrix}.
    \end{align*}
    The diagonal elements of the diagonal matrix $\mathbf{\Lambda}$ correspond to the eigenvalues of $\mathbf{G}^{\parallel}_{\text{t}}$ since the Kronecker product between two orthogonal matrices are orthogonal, i.e., $\left(\mathbf{Q}_{\text{v}} \otimes \mathbf{Q}_{\text{h}}\right)\left(\mathbf{Q}_{\text{v}} \otimes \mathbf{Q}_{\text{h}}\right)^{*} = \mathbf{I}$. The direct expansion of $\mathbf{\Lambda}$ finishes the proof.
\end{IEEEproof} \vspace{0.1in}

Based on Corollary \ref{thm:cor1}, we can determine the optimal spacing $\Delta_{\text{v}}$ and $\Delta_{\text{h}}$. To maximize spectral efficiency \eqref{eqn:P2}, antenna spacing should satisfy $4\left\lfloor\frac{\Delta_{\text{v}}}{2} N_{\text{v, min}}\right\rfloor \left\lfloor\frac{\Delta_{\text{h}}}{2} N_{\text{h, min}}\right\rfloor = N_s$. Let us consider the decomposition of $N_s$ as $N_s=N_{s,\text{v}}N_{s,\text{h}}$ to derive the optimal antenna configuration. Then, the antenna spacing that satisfies
\begin{align} \label{eqn:Nsdtdr}
    2\left\lfloor\frac{d_{t,i}d_{r,i}N_{i}M_{i}}{2\lambda_{c} D}\right\rfloor = N_{s,i} \quad \forall i \in \{\text{v, h}\}
\end{align}
maximizes the spectral efficiency \eqref{eqn:P2}. Note that $N_s$ can always be expressed as the product of two integers $N_{s,\text{v}}$ and $N_{s,\text{h}}$ but it is not unique since the integer product is not unique. The proposed antenna configuration, under the paraxial propagation condition, in conjunction with the upper bound in \eqref{eqn:maxmaxSE}, highlights a significant aspect in point-to-point (p2p) LoS MIMO systems. In such systems, where the transceiver has a limited number of RF chains, the spatial multiplexing gain primarily depends on the aperture size, regardless of the underlying array antenna type, whether phase-shifter or metasurface. This observation aligns with prior research \cite{di2023mimo}, which demonstrated that the  DoF of a fully-digital p2p LoS MIMO system are determined by the size of the arrays' aperture.

\begin{figure}[t!]
    \centering
    \includegraphics[scale=0.47]{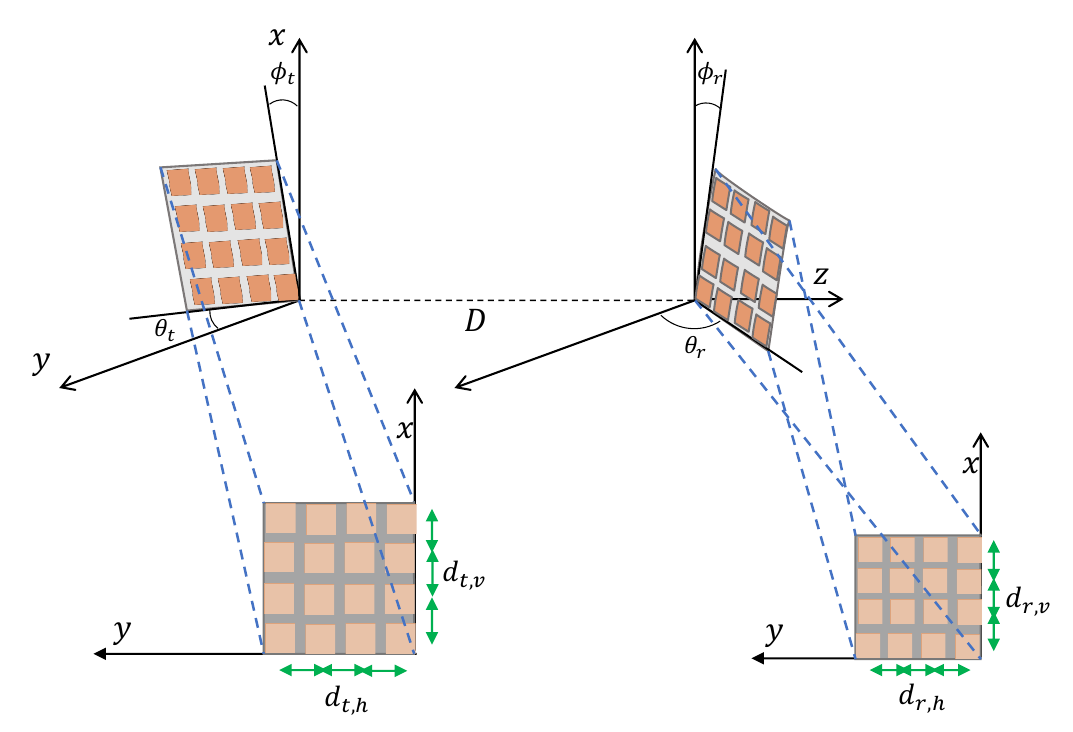}
    \caption{Examples of optimally spaced planar array with arbitrary rotation.} 
    \label{fig:genericopt}
\end{figure}

To figure out the optimum antenna configuration for non-parallel, arbitrary rotated UPAs, we leverage the fact that the spectral efficiency of the channel $\tilde{\mathbf{H}}$ is invariant under phase offsets along the link direction. Specifically, phase offsets along the link direction are reflected in $\mathbf{D}_{\text{r}}$ and $\mathbf{D}_{\text{t}}$, which do not change the eigenvalues of $\tilde{\mathbf{H}}$, and thus do not affect $\mathbf{H}$ as well. Since the upper bound of spectral efficiency is an additively separable function of the eigenvalues, the maximum spectral efficiency is achieved when the matrix $\tilde{\mathbf{H}}$ of the non-parallel system coincides with the parallel counterpart. That is, for the generic planar array system, the uniform parallelogram array such that the $xy$-plane coordinate of antennas remain the same as the optimal antenna arrangement in parallel UPAs while the $z$-axis coordinate of antennas change with the 3D rotation angles $\theta_\text{t}, \phi_\text{t}, \theta_\text{r}, \phi_\text{r}$ maximizes the spectral efficiency \eqref{eqn:P2}, which is depicted in Fig. \ref{fig:genericopt}. To present the main result, let $m_{\text{v}} \triangleq \left\lfloor \frac{m}{\sqrt{M_{\text{h}}}} \right\rfloor$ and $m_{\text{h}} \triangleq m ~(\mod \sqrt{M_{\text{h}}})$ be the location variables corresponding to the $m$-th element in the Tx planar array. Also, let $n_{\text{v}} \triangleq \left\lfloor \frac{n}{\sqrt{N_{\text{h}}}} \right\rfloor$ and $n_{\text{h}} \triangleq n ~(\mod \sqrt{N_{\text{h}}})$ be the location variables corresponding to the $n$-th element in the Rx planar array. The optimum antenna configuration for the arbitrary rotated UPA is given as in the following theorem.

\begin{theorem} \label{thm:optimal_sapcing}
  The optimum antenna configuration for the arbitrary rotated UPA equipped with hybrid architecture in Fig. \ref{fig:systemmodel} is obtained as
\begin{align} \label{eqn:optnonpara}
\begin{split}
    &\begin{pmatrix}
        \mathrm{t}^{x}_{m} \\
        \mathrm{t}^{y}_{m} \\
        \mathrm{t}^{z}_{m} 
    \end{pmatrix}
    =
    \begin{pmatrix}
        d_{\mathrm{t,v}} m_{\mathrm{v}} \\
        d_{\mathrm{t,h}} m_{\mathrm{h}} \\
        \frac{\cos\theta_{\mathrm{t}} \sin \phi_{\mathrm{t}} d_{\mathrm{t,v}} m_{\mathrm{v}} + \sin \theta_{\mathrm{t}} d_{\mathrm{t,h}} m_{\mathrm{h}}}{-\cos \theta_{\mathrm{t}} \cos \phi_{\mathrm{t}}}
    \end{pmatrix},
    \\
    &\begin{pmatrix}
        \mathrm{r}^{x}_{n}\\
        \mathrm{r}^{y}_{n}\\
        D+\mathrm{r}^{z}_{n}
    \end{pmatrix}
    =
    \begin{pmatrix}
        d_{\mathrm{r,v}} n_{\mathrm{v}} \\
        d_{\mathrm{r,h}} n_{\mathrm{h}} \\
        D+\frac{\cos\theta_{\mathrm{r}} \sin \phi_{\mathrm{r}} d_{\mathrm{r,v}} n_{\mathrm{v}} + \sin \theta_{\mathrm{r}} d_{\mathrm{r,h}} n_{\mathrm{h}}}{\cos \theta_{\mathrm{r}} \cos \phi_{\mathrm{r}}}
    \end{pmatrix}.
\end{split}
\end{align} 
where the antenna spacing $d_{\mathrm{t,v}}, d_{\mathrm{t,h}}, d_{\mathrm{r,v}}$, and $d_{\mathrm{r,h}}$ satisfy
\begin{align*}
    2\left\lfloor\frac{d_{\mathrm{t},i}d_{\mathrm{r},i}N_{i}M_{i}}{2\lambda_{c} D}\right\rfloor = N_{s,i} \quad \forall i \in \{\mathrm{v, h}\}
\end{align*}
depending on the number of data stream $N_s$.
\end{theorem}

\begin{IEEEproof}
 See Appendix \ref{proof1}.
\end{IEEEproof}\vspace{0.1in}

Theorem \ref{thm:optimal_sapcing} reveals that when antenna arrays are perpendicular to the communication link, the optimal configuration is the uniform parallelogram rather than the uniform rectangular array. Note that the above antenna configuration reduces to the optimal configuration of the parallel UPA setup with $\theta_t=\phi_t=\theta_r=\phi_r=0$, that is, when $\text{t}^{z}_{m}$ and $\text{r}^{z}_{n}$ reduce to $0$. 

\begin{remark}
In practical communication systems where both communication distances and array rotations vary dynamically, the adoption of antenna selection techniques offers advantages over physically adjusting the location of antenna elements. Such techniques enable the dynamic selection of a subset of available antenna elements from a predetermined array by electronically turning them on or off, facilitating the implementation of optimal inter-antenna spacing. This optimal array configuration can be also achieved through the utilization of fluid antenna systems \cite{wong2020fluid}, where position-flexible arrays allow for the movement of antenna elements to desirable locations. Furthermore, a fixed array configuration can be utilized to minimize capacity fluctuation over a range of distances. This can be achieved by integrating the proposed array configuration for hybrid MIMO systems with the robust array configuration as proposed in \cite{palaiologos2023non}.
\end{remark}

\begin{figure}[t!]
    \centering
    \includegraphics[scale=0.55]{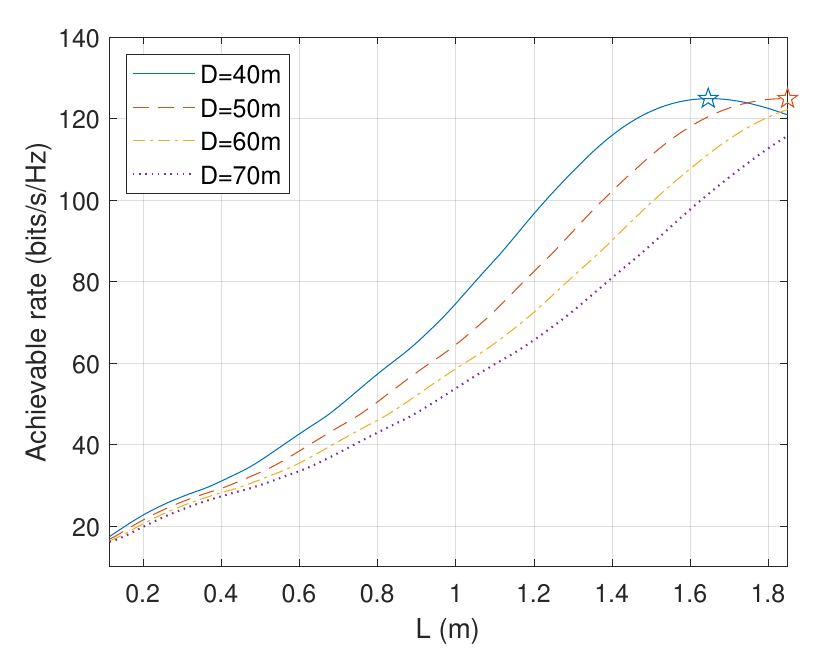}
    \caption{Achievable rate on the array aperture for the different communication distance with the number of antenna elements $N_t = N_r = 256$, the number of data stream and RF chain $M_f=N_f=N_s=16$. The pentagram marker indicates the array aperture of the optimally spaced array.} 
    \label{fig:limaperture}
\end{figure}

When considering the limitations imposed on the aperture of both the transmitter and receiver arrays, achieving the optimal configuration of the antenna array may become unattainable. Specifically, the aperture of the array must adhere to the inequality given as $$
L_{\text{t}} L_{\text{r}} \ge 2\sqrt{N_{s}} \lambda_{c} D$$
in order to reach the upper bound as described in \eqref{eqn:maxmaxSE}. The detailed derivation of the aperture constraint is delegated in Appendix \ref{aperture_proof}. In the cases where the system is constrained by the aperture of the array, with $L_{\text{t}} L_{\text{r}}$ strictly less than $ 2\sqrt{N_{s}} \lambda_{c} D$, optimal deployment necessitates maximizing the spatial separation between adjacent antennas within the confined aperture as depicted in Fig. \ref{fig:limaperture}. 

\section{Analog-Digital Beam Focusing Design with Given Antenna Configuration under Limited Antenna Aperture } \label{beamforming}

 The development of a hardware-constrained precoder and combiner is essential to achieve the spatial multiplexing gain of the optimally configured MIMO system. Even with antenna arrays optimally configured at both the transmitter and receiver, achieving higher data rates is not possible without implementing suitable precoders and combiners. To our knowledge, the joint design of a hybrid precoder and combiner in the near-field region remains unexplored. While DMA-based hybrid transceivers have received recent attention in research, our finding in Section \ref{optspacing} indicate that such architecture does not offer an advantage over phase-shifter-based hybrid architecture in terms of data rate in the considered system, although they may have advantages in terms of hardware, such as power consumption. Additionally, DMA-based transceivers utilize sub-wavelength spaced elements, which are unfavorable in LoS MIMO systems where a large aperture size is necessary to achieve high data rates. Therefore, in this section, our focus is on exploring near-field beam focusing via phase-shifter-based hybrid precoder and combiner that enables the achievement of high data rates in LoS MIMO systems.

In particular, the phase-shifters based hybrid transmitter uses a precoder $\mathbf{F}=\mathbf{F}_{\text{RF}}\mathbf{F}_{\text{BB}}$, where $\mathbf{F}_{\text{BB}} \in \mathbb{C}^{M_{\text{f}} \times N_s}$ is a baseband precoder and $\mathbf{F}_{\text{RF}} \in \mathbb{C}^{M \times M_{\text{f}}}$ is an analog precoder. The analog precoder $\mathbf{F}_{\text{RF}}$ is implemented using phase-shifters, ensuring that all its elements have an equal norm. Similarly, the receiver receives $N_s$ data streams with $N_{\text{f}}$ RF chains, satisfying $N_s \le N_{\text{f}} \le N$. The corresponding combiner has the form $\mathbf{W}=\mathbf{W}_{\text{RF}}\mathbf{W}_{\text{BB}}$ with a digital combiner $\mathbf{W}_{\text{BB}}\in\mathbb{C}^{N_{\text{f}} \times N_s}$ and an analog combiner $\mathbf{W}_{\text{RF}} \in \mathbb{C}^{N \times N_{\text{f}}}$ with an equal norm constraint. Designing a hybrid antenna architecture for an array configured near optimally involves addressing the optimization problem:
\begin{align} \label{eqn:noncvx}
\begin{split}
    (\mathbf{F}_{\text{RF}}, \mathbf{F}_{\text{BB}}, &\mathbf{W}_{\text{RF}}, \mathbf{W}_{\text{BB}}) =\argmax_{\mathbf{F}_{\text{RF}}, \mathbf{F}_{\text{BB}}, \mathbf{W}_{\text{RF}}, \mathbf{W}_{\text{BB}}} R \\
    &\text{s.t.} ~~ \text{Tr} \left( \mathbf{F}^{*} \mathbf{F} \right) \le P_{\text{t}},\\
    &\quad  ~~ \textbf{F}_{\text{RF}} \in \mathcal{F}_{\text{RF}}, \\
    &\quad ~~ \mathbf{W}_{\text{RF}} \in \mathcal{W}_{\text{RF}},
\end{split}
\end{align} 
under the Gaussian signaling assumption with $\mathbf{x}=\mathbf{F}\mathbf{s}$ where $\mathbf{s}$ is the length $N_s$ vector consists of unit-power symbols, $\mathcal{F}_{\text{RF}}$ is the set of feasible RF precoders, $\mathcal{W}_{\text{RF}}$ is the set of feasible RF combiners and $P_{\text{t}}$ is a transmit power constraint. The problem \eqref{eqn:noncvx} is a non-convex problem due to the constant modulus constraint and the coupled optimization variables. Thus, rather than directly solving the non-convex problem, the reasonable approximation is to minimize the distance between the hybrid solution and the unconstrained solution which separates the coupled optimization variables. In particular, minimizing the distance between two matrices in the Frobenius norm sense has been widely used in the hybrid beamforming literature \cite{zhang2022beam, el2014spatially, alkhateeb2014mimo, sohrabi2016hybrid}. 

Let   $\mathbf{F}_{\text{opt}}=[\mathbf{v}_1, \ldots, \mathbf{v}_{N_s}]$ and $\mathbf{W}_{\text{opt}}=[\mathbf{u}_1, \ldots, \mathbf{u}_{N_s}]$ be the optimal fully-digital precoder and combiner to transmit $N_s$ data streams. Then,  the resulting approximated hybrid precoder optimization problem  is reformulated as
\begin{align} \label{eqn:hybridF}
\begin{split}
    (\mathbf{F}_{\text{RF}}, \mathbf{F}_{\text{BB}}) &= \argmin_{\mathbf{F}_{\text{RF}},\mathbf{F}_{\text{BB}}} \lVert \mathbf{F}_{\text{opt}} - \mathbf{F}_{\text{RF}}\mathbf{F}_{\text{BB}} \rVert_{F},\\
    &\text{s.t.} \quad \mathbf{F}_{\text{RF}} \in \mathcal{F}_{\text{RF}}, \\
    &\quad \quad  \text{Tr} (\mathbf{F} \mathbf{F}^{*}) \le P_{\text{T}},
\end{split}
\end{align}
and the corresponding hybrid combiner optimization problem is given by
\begin{align} \label{eqn:hybridW}
\begin{split}
    (\mathbf{W}_{\text{RF}}, \mathbf{W}_{\text{BB}}) = &\argmin_{\mathbf{W}_{\text{RF}},\mathbf{W}_{\text{BB}}} \lVert \mathbf{W}_{\text{opt}} - \mathbf{W}_{\text{RF}}\mathbf{W}_{\text{BB}} \rVert_{F},\\
    &\text{s.t.} \quad \mathbf{W}_{\text{RF}} \in \mathcal{W}_{\text{RF}}.
\end{split}
\end{align}
Unfortunately, the optimization problems specified in \eqref{eqn:hybridF} and \eqref{eqn:hybridW} persist as non-convex due to the inclusion of non-convex constraints. Consequently, the task of determining the optimal hybrid precoder and combiner, intended to minimize the gap between the hybrid and unconstrained precoder/combiner and thereby enhance spectral efficiency, proves to be formidable.

 To provide a tractable design of precoder and combiner while avoiding the deployment of the computationally demanding optimization method, we examine the asymptotic eigenvector spaces of the LoS MIMO channel. We capitalize on the relationship \eqref{eqn:SVD} between singular vectors of $\mathbf{H}$ and $\tilde{\mathbf{H}}$.  Additionally, we leverage the fact that the left singular vectors of $\tilde{\mathbf{H}}$ coincide with the eigenvectors of $\mathbf{G}_{\text{r}}$ and the right singular vectors of $\tilde{\mathbf{H}}$ coincide with the eigenvectors of $\mathbf{G}_{\text{t}}$ where the channel gain matrices are defined in Section \ref{optspacing}. The eigenvector spaces of the asymptotic channel gain matrices have deterministic characteristics, which are stated in the following theorem.

\begin{theorem} \label{thm:Hgain}
For the generic UPA wide-aperture MIMO, the channel gain matrices $\mathbf{G}_{\mathrm{r}}$ and $\mathbf{G}_{\mathrm{t}}$ are a Hermitian doubly block Toeplitz matrix and it can be asymptotically diagonalized by the 2D-DFT matrix as
\begin{align}
    \mathbf{G}_{\mathrm{r}} = \left( \mathbf{\Omega}_{N_{\mathrm{v}}} \otimes \mathbf{\Omega}_{N_{\mathrm{h}}} \right) \mathbf{\Lambda} \left( \mathbf{\Omega}_{N_{\mathrm{v}}} \otimes \mathbf{\Omega}_{N_{\mathrm{h}}} \right)^{*}, \\
    \mathbf{G}_{\mathrm{t}} = \left( \mathbf{\Omega}_{M_{\mathrm{v}}} \otimes \mathbf{\Omega}_{M_{\mathrm{h}}} \right) \mathbf{\Lambda} \left( \mathbf{\Omega}_{M_{\mathrm{v}}} \otimes \mathbf{\Omega}_{M_{\mathrm{h}}} \right)^{*},
\end{align}
where $\mathbf{\Omega}_{k}$ is a k dimension discrete Fourier transform (DFT) matrix and $\mathbf{\Lambda}$ is the diagonal eigenvalue matrix. 
\end{theorem}
\begin{IEEEproof}
See Appendix \ref{proof2}.
\end{IEEEproof}\vspace{0.1in}

Theorem \ref{thm:Hgain} establishes that the asymptotic channel gain matrix can be expressed as a linear combination of 2D-DFT vectors. This finding extends the representation of the channel gain matrix in ULA wide-aperture MIMO from a Toeplitz matrix to a doubly block Toeplitz matrix. Building upon Theorem \ref{thm:Hgain} and  \eqref{eqn:SVD}, we derive the asymptotic right and left singular vector spaces of the original right singular space $\mathbf{V}$ and the  original left singular vector space $\mathbf{U}$ of the channel matrix $\mathbf{H}$, respectively, as
\begin{align}\label{eqn:singularvectors}
\begin{split}
    \tilde{\mathbf{V}}_{\text{M}} = \mathbf{D}^{*}_{t} \left( \mathbf{\Omega}_{M_{\text{v}}} \otimes \mathbf{\Omega}_{M_{\text{h}}} \right)^{*} ~\in   \mathbb{C}^{M \times M}, \\
    \tilde{\mathbf{U}}_{\text{N}} = \mathbf{D}^{*}_r \left( \mathbf{\Omega}_{N_{\text{v}}} \otimes \mathbf{\Omega}_{N_{\text{h}}} \right)^{*} ~ \in \mathbb{C}^{N \times N}.
\end{split}
\end{align}

As the number of antenna increases, the right singular vectors of $\mathbf{H}$ converge to $\tilde{\mathbf{V}}_{\text{M}}$ and the left singular vectors of $\mathbf{H}$ converge to $\tilde{\mathbf{U}}_{\text{N}}$. Since the multiplication of two unitary phase factors is unitary, the elements of such matrices have an equal norm and $\tilde{\mathbf{V}}_{\text{M}}$ and $\tilde{\mathbf{U}}_{\text{N}}$ are unitary matrices. Therefore, we have the asymptotically optimal hybrid precoder $\mathbf{F}_{\text{RF}} = [\tilde{\mathbf{V}}_{\text{M}}]_{1:M_{f}} \in \mathbb{C}^{M \times M_{\text{f}}}$, $\mathbf{F}_{\text{BB}}=[\frac{1}{N_s}\mathbf{I}_{N_s} \parallel \mathbf{0}]^{T} \in \mathbb{C}^{M_{\text{f}} \times N_s}$ and the corresponding hybrid combiner $\mathbf{W}_{\text{RF}} = [\tilde{\mathbf{U}}_{\text{N}}]_{1:N_{f}} \in \mathbb{C}^{N \times N_{\text{f}}}$, $\mathbf{W}_{\text{BB}}=[\mathbf{I}_{N_s} \parallel \mathbf{0}]^{T} \in \mathbb{C}^{N_{\text{f}} \times N_s}$. 

In fact, for large $M$ and $N$, the last $M_f - N_s$ columns of the analog precoder $\mathbf{F}_{\text{RF}}$ can take on any values since those columns are multiplied by zeros. Similarly, the last $N_f - N_s$ columns of the analog combiner can consist of any feasible candidates. Therefore, utilizing the minimum number of RF chains $M_{\text{f}} = N_{\text{f}}=N_s$ is sufficient to achieve the spectral efficiency attained by the unconstrained precoder and combiner. 

\begin{corollary} \label{thm:asyBF}
For asymptotically large $M$ and $N$, the optimal hybrid precoder and the optimal hybrid combiner are obtained, respectively, as
    \begin{align}
    \mathbf{F}_{\mathrm{RF}} = [\mathbf{D}^{*}_t \left( \mathbf{\Omega}_{M_{\mathrm{v}}} \otimes \mathbf{\Omega}_{M_{\mathrm{h}}} \right)^{*}]_{1:N_{\mathrm{s}}}, \mathbf{F}_{\mathrm{BB}}=\frac{1}{N_s}\mathbf{I}_{N_s} \\
    \mathbf{W}_{\mathrm{RF}} = [\mathbf{D}^{*}_r \left( \mathbf{\Omega}_{N_{\mathrm{v}}} \otimes \mathbf{\Omega}_{N_{\mathrm{h}}} \right)^{*}]_{1:N_{\mathrm{s}}}, \mathbf{W}_{\mathrm{BB}}=\mathbf{I}_{N_s}.
    \end{align}
\end{corollary}
\begin{IEEEproof}
It is a direct consequence of Theorem \ref{thm:Hgain}.
\end{IEEEproof}\vspace{0.1in}

For moderate $M$ and $N$,  the structure of the LoS wide-aperture MIMO channel allows for an approximate low-dimensional change of basis. We aim to leverage the inherent characteristics of the LoS wide-aperture MIMO channel, which can be expressed as a linear combination of weighted 2D-DFT vectors as deduced from Theorem \ref{thm:Hgain}. Specifically, we can construct $\mathbf{F}_{\text{RF}}$ as $M_{\text{f}}$ columns of $\tilde{\mathbf{V}}_{\text{M}}$ that serve as the optimal low-dimensional representation of $\mathbf{F}_{\text{opt}}$. Simultaneously, we can build $\mathbf{F}_{\text{BB}}$ to express the appropriate linear combination that transforms the space of $\mathbf{F}_{\text{RF}}$ into the space of $\mathbf{F}_{\text{opt}}$. The determination of the best low-dimensional representation can be addressed through matching pursuit algorithms that provide a solution to the following optimization problem: 
 
 \begin{align} \label{eqn:sparseF}
\begin{split}
    (\hat{\mathbf{F}}_{\text{BB}}) = &\argmin_{\hat{\mathbf{F}}_{\text{BB}}} \lVert \mathbf{F}_{\text{opt}} - \tilde{\mathbf{V}}_{\text{M}} \hat{\mathbf{F}}_{\text{BB}} \rVert_{F},\\
\end{split}
\end{align}
where $\hat{\mathbf{F}}_{\text{BB}} \in \mathbb{C}^{M \times N_s}$ is a row-sparse matrix that has $M_{\text{f}}$ non-zero rows. As we solve the sparse recovery problem in \eqref{eqn:sparseF}, the baseband precoder $\mathbf{F}_{\text{BB}}$ can be obtained as the $M_{\text{f}}$ non-zero rows of $\hat{\mathbf{F}}_{\text{BB}}$ and the analog precoder $\mathbf{F}_{\text{RF}}$ can be obtained as the corresponding $M_{\text{f}}$ columns of $\tilde{\mathbf{V}}_{\text{M}}$. It is worth noting that although the underlying physics of the channel is fundamentally different from \cite{el2014spatially}, the resulting optimization problem is of similar form. 

The sparse reconstruction algorithm iteratively estimates the support of $\tilde{\mathbf{V}}_{\text{M}}$ and updates the residual by removing the effect of previously estimated supports. Among several sparse recovery algorithms, we adopt orthogonal matching pursuit(OMP) to solve \eqref{eqn:sparseF}. The OMP at the $i$-th iteration is defined as follows:
\begin{align}
    k = \argmax_{k} &\left(\left[
    \left(\tilde{\mathbf{V}}_{\text{M}}^{*} \mathbf{F}^{i-1}_{\text{res}} \right) \left(\tilde{\mathbf{V}}_{\text{M}}^{*} \mathbf{F}^{i-1}_{\text{res}}\right)^{*}
    \right]_{k,k}\right), \\
    &\mathbf{F}^{i}_{\text{RF}} = \left[ \mathbf{F}^{i-1}_{\text{RF}} \parallel
    \left[ \tilde{\mathbf{V}}_{\text{M}} \right]_{:,k} \right],\\
    &\mathbf{F}^{i}_{\text{BB}} =  \left( (\mathbf{F}^{i}_{\text{RF}})^* \mathbf{F}^{i}_{\text{RF}} \right)^{-1} 
    \left( \mathbf{F}^{i}_{\text{RF}} \right)^{*} \mathbf{F}_{\text{opt}}, \label{eqn:LSsolution}\\
    &\mathbf{F}^{i}_{\text{res}} = \frac{\mathbf{F}_{\text{opt}}- \mathbf{F}^{i}_{\text{RF}}\mathbf{F}^{i}_{\text{BB}} }{\lVert \mathbf{F}_{\text{opt}}- \mathbf{F}^{i}_{\text{RF}}\mathbf{F}^{i}_{\text{BB}} \rVert^{2}_{F}}
\end{align}
with initial values $\mathbf{F}^{0}_{\text{RF}} = [\,]$ and $\mathbf{F}^{0}_{\text{res}} = \mathbf{F}_{\text{opt}}$, where $[\,]$ denotes an empty set. The baseband precoder of the $i$-th iteration is obtained as the least-squares solution of \eqref{eqn:sparseF}.  For each iteration, the OMP algorithm identifies the most suitable vector by seeking the maximum projection. This involves incorporating the best-matching column of $\tilde{\mathbf{V}}_{\text{M}}$ into the analog precoder $\mathbf{F}_{\text{RF}}$. Subsequently, the algorithm obtains the corresponding baseband precoder $\mathbf{F}_{\text{BB}}$ through a least-squares approach, eliminating the contribution of the already selected support of $\tilde{\mathbf{V}}_{\text{M}}$ by computing the residual space. The process then continues to identify the next best-matching column of $\tilde{\mathbf{V}}_{\text{M}}^{*}$. It is noteworthy that the OMP algorithm requires $M_{\text{f}}$ iterations to achieve the desired hybrid precoder.

\begin{figure}[t!]
    \centering
    \includegraphics[scale=0.6]{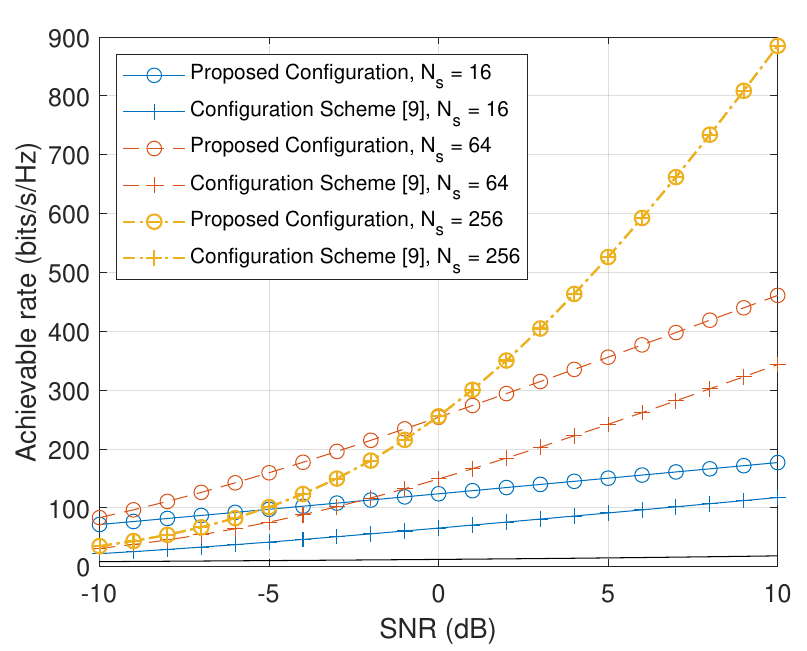}
    \caption{Upper bound of the achievable rate for the different array configuration on the SNR.}
    \label{fig:spacing}
\end{figure}

For the hybrid combiner design in the LoS wide-aperture MIMO channel, we can approximate $\mathbf{W}_{\text{RF}}$ to be $N_{\text{f}}$ columns of $\tilde{\mathbf{U}}_{\text{N}}$ that are the best low-dimensional representation of $\mathbf{W}_{\text{opt}}$ and construct $\mathbf{W}_{\text{BB}}$ to be an appropriate linear combination that transforms the space of $\mathbf{W}_{\text{RF}}$ into the space of $\mathbf{W}_{\text{opt}}$ by exploiting the structure of the LoS wide-aperture MIMO channel. As a consequence, the non-convex problem in \eqref{eqn:hybridW} can be replaced by
\begin{align} \label{eqn:sparseW}
    (\hat{\mathbf{W}}_{\text{BB}}) = &\argmin_{\hat{\mathbf{W}}_{\text{BB}}} \lVert \mathbf{W}_{\text{opt}} - \tilde{\mathbf{U}}_{\text{N}} \hat{\mathbf{W}}_{\text{BB}} \rVert_{F},
\end{align}
where $\hat{\mathbf{W}}_{\text{BB}} \in \mathbb{C}^{N \times N_s}$ is a $N_{\text{f}}$ row-sparse matrix. The hybrid precoder design in \eqref{eqn:sparseF} and the hybrid combiner design in \eqref{eqn:sparseW} are essentially equivalent in form to the LoS wide-aperture MIMO channel. To derive the hybrid combiner, we employ the OMP algorithm. Specifically, we initialize the values as follows: $\mathbf{W}^{0}_{\text{RF}} = [\,]$, $\mathbf{W}^{0}_{\text{res}} = \mathbf{W}_{\text{opt}}$. Subsequently, we perform $N_{\text{f}}$ iterations of the OMP algorithm to iteratively refine and obtain the desired hybrid combiner.

\section{Numerical results} \label{results}
In this section, we evaluate the achievable rate for the proposed antenna configuration as detailed in Section \ref{optspacing}. We then compare this configuration with the existing UPA antenna configuration. Furthermore, we analyze the performance of the proposed  analog-digital beam focusing scheme, elaborated in Section \ref{beamforming}. Throughout the simulations, we maintained a carrier frequency of $28$ GHz and a communication distance $D=50$ m.  We also ensure an equal number of   RF chains at both the transmitter and receiver, with the number of RF chains matching the number of transmitting data streams, i.e., $M_{\text{f}}=N_{\text{f}}=N_{s}$. Moreover, unless otherwise specified, we utilized the optimally configured array as discussed in Section \ref{optspacing}.

\begin{figure}[t!]
    \centering
    \includegraphics[scale=0.6]{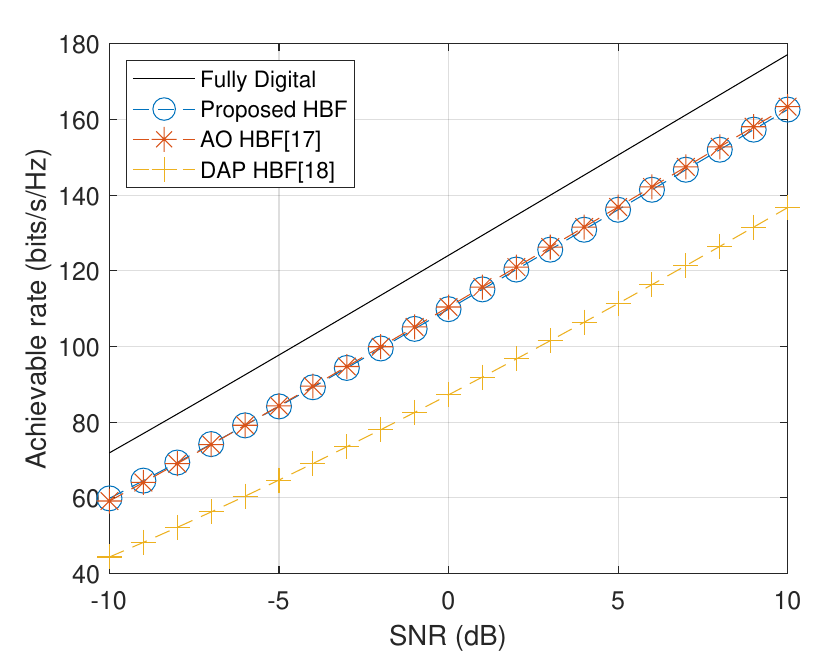}
    \caption{Achievable rate of the different analog-digital hybrid beam focusing schemes on the SNR.}
    \label{fig:SNR}
\end{figure}

Fig. \ref{fig:spacing} illustrates the upper bound of the achievable rate for the proposed array configuration discussed in Section \ref{optspacing}, employing a square array with $N=M=256$ antennas. This upper bound represents the fully digital beamforming rate utilizing the precoder $\mathbf{F}_{\text{opt}}$ and the combiner $\mathbf{W}_{\text{opt}}$ in \eqref{eqn:hybridF} and \eqref{eqn:hybridW}, respectively. The solid black line represents the achievable rate for the conventional $\lambda/2$-spaced antenna array. Compared to the planar array configuration scheme presented in \cite{song2015spatial}, our proposed array configuration scheme demonstrates superior performance in terms of data rate.  It is evident that communicating with a larger number of data streams yields inferior results in low SNR regimes, despite sufficient RF chains at both the transmitter and receiver ends. However, as SNR increases, the data rate significantly improves. The impact of SNR on the maximum achievable rate for the fully digital configuration was previously investigated in \cite{do2020reconfigurable}. Particularly in low-power regimes, the optimal strategy involves concentrating channel gain into a smaller number of sub-channels. We have observed that this property holds true for the hybrid MIMO architecture as well.

\begin{figure}[t!]
    \centering
    \includegraphics[scale=0.6]{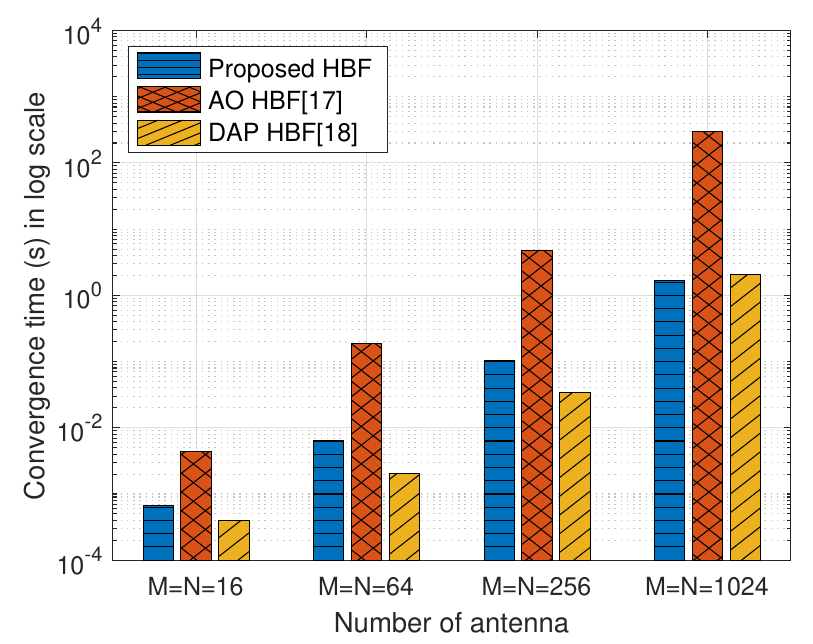}
    \caption{Average CPU time versus the number of antennas.}
    \label{fig:convergence_time}
\end{figure}

\begin{figure}[t!]
    \centering
    \includegraphics[scale=0.6]{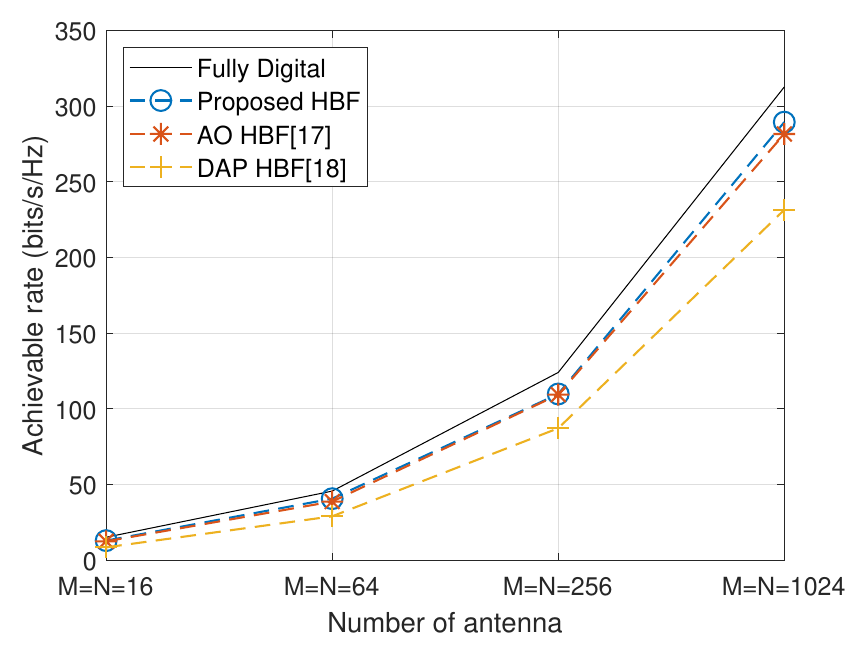}
    \caption{Achievable rate of the different hybrid beamfomring schemes in the number of antennas.}
    \label{fig:convergence_time_SE}
\end{figure}

 Fig. \ref{fig:SNR} shows the achievable rates for various beam focusing schemes employing analog-digital architecture. We conducted simulations using a square array comprising $N=M=256$ antennas, with $N_s=16$ and SNR= $0$ dB. The achievable rate of fully digital beamforming serves as an upper bound for comparison.  We observe that the performance of the proposed hybrid analog-digital beam focusing closely follows the performance upper bound across a range of SNR values, from low to high. The beam focusing scheme in \cite{zhang2022beam} uses alternating optimization algorithm that alternately optimizes the analog beamforming matrix and the digital beamforming matrix until convergence. The alternating optimization algorithm closely tracks the optimal precoder and the combiner, but the computational complexity is high due to the need for numerous matrix inversion computations.  In \cite{wu2022distance}, the phase of the optimal beamforming matrix as the analog beamforming matrix, while the digital beamforming matrix is obtained based on the SVD of the effective channel with the dynamic connection between the RF chains and the antenna elements. As depicted in Fig. \ref{fig:SNR}, the proposed beam focusing scheme significantly outperforms \cite{wu2022distance}.

Fig. \ref{fig:convergence_time} illustrates the average CPU time versus the number of antennas for different analog-digital beam focusing schemes.  The CPU time measurements were obtained using an Intel Xeon Silver 4210R processor through Monte Carlo simulations to derive the average runtime. Furthermore, Fig. \ref{fig:convergence_time_SE} illustrates the achievable rate performance for different analog-digital beam focusing schemes. For both sets of results, we consider parallel transmission and reception with the number of data streams $N_s = \sqrt{N}$ and an SNR of $0$ dB. Both figures highlight that our proposed scheme exhibits significantly lower computational complexity compared to the approach presented in \cite{zhang2022beam}, while maintaining closely aligned achievable rates with optimal performance. Furthermore, while the computational complexity of the scheme proposed in \cite{wu2022distance} is lower than ours for a small number of antennas, it becomes slightly larger as the number of antennas increases.

\begin{figure}[t!]
    \centering
    \includegraphics[scale=0.6]{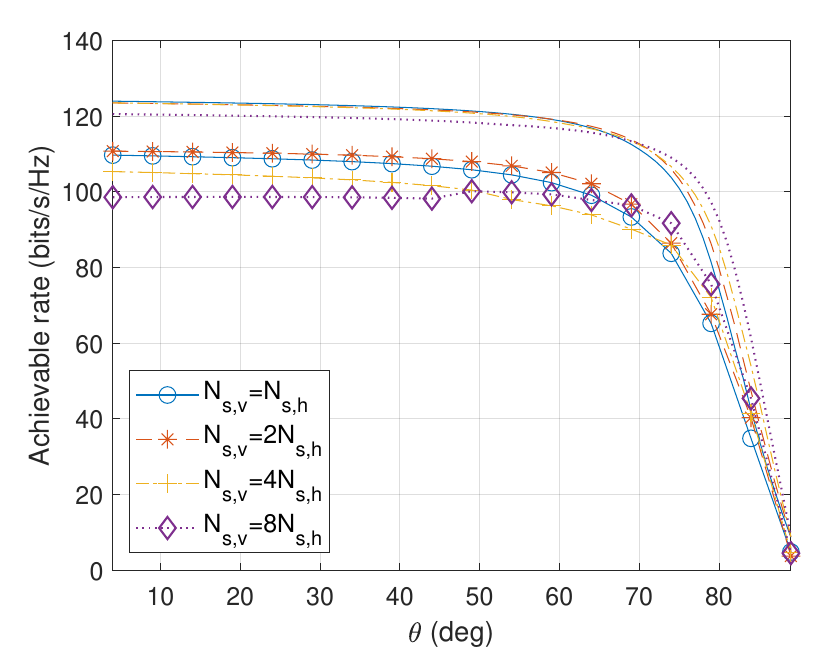}
    \caption{Achievable rate of the proposed analog-digital hybrid beam focusing scheme on different 3D rotation angle $\theta$.}
    \label{fig:diff_Ns_sqaure}
\end{figure}

\begin{figure}[t!]
    \centering
    \includegraphics[scale=0.6]{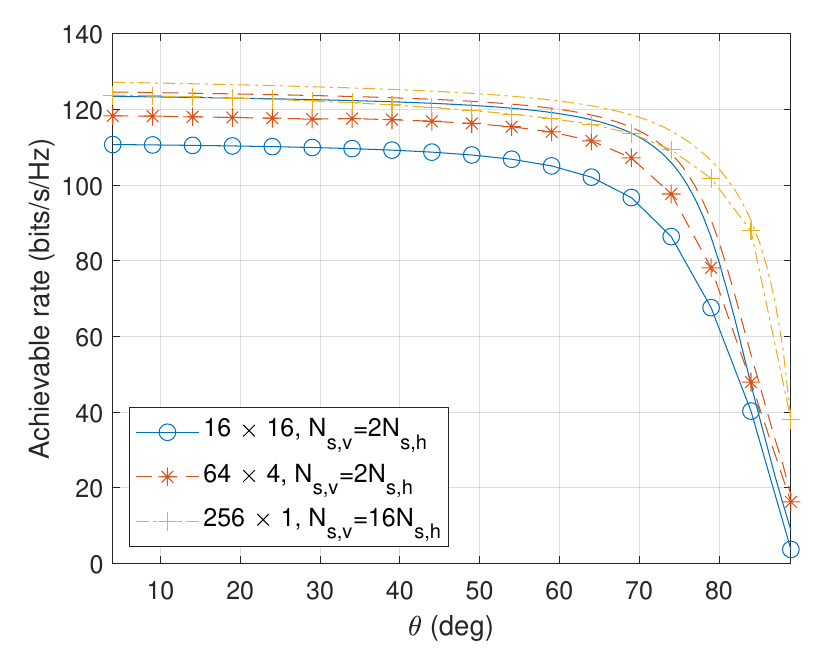}
    \caption{Achievable rate of the proposed analog-digital hybrid beam focusing scheme for different rectangular arrays on different 3D rotation angle $\theta$.}
    \label{fig:diff_Ns_n_shape}
\end{figure}

Fig. \ref{fig:diff_Ns_sqaure} demonstrates the achievable rate of the proposed analog-digital beam focusing scheme for $0$ dB SNR at various 3D rotation angles.  The line without markers depicts the performance of fully-digital transmitters and receivers with a limited number of data streams, serving as an upper bound for performance comparison. For the illustration, we set $\theta_t=\phi_t=\theta_r=\phi_r$ as $\theta$. Different tuples of data streams ${N_{s,\text{v}}, N_{s,\text{h}}}$ were considered while maintaining $N_s=16$ and $N=M=256$. The solid line represents the upper limit of the rate. As shown in Fig. \ref{fig:diff_Ns_n_shape}, the proposed beam focusing scheme achieves near-optimal data rates across a wide range of rotation angles. The degradation of the achievable rate at the higher degrees of tilt is attributed to the effective antenna array transforming into linear arrays with a reduced number of antenna elements in highly tilted scenarios. Although the choice of different $N_{s,\text{v}}, N_{s,\text{h}}$ does not theoretically affect the achievable rate, the variation in the choice of $N_{s,\text{v}}, N_{s,\text{h}}$ alters the array aperture, leading to a degradation in the accuracy of the approximation given in \eqref{eqn:approxH}. Consequently, the performance of the proposed analog-digital beam focusing scheme varies with different choices of $N_{s,\text{v}}, N_{s,\text{h}}$, as depicted in Fig. \ref{fig:diff_Ns_sqaure}.

For Fig. \ref{fig:diff_Ns_n_shape}, we considered the rectangular Tx and Rx with total $256$ antenna elements and evaluated the achievable rate at different 3D rotation angles. Also, we considered different $N_{s,\text{v}}, N_{s,\text{h}}$ while keeping the number of data streams $N_s$ same. The line without marker represents the performance of fully digital Tx and Rx with a limited number of data streams, which serves as an upper bound of the performance. As shown in Fig. \ref{fig:diff_Ns_n_shape}, the proposed analog-digital beam focusing scheme closely tracks the optimal performance. 
In Fig. \ref{fig:diff_Ns_n_shape}, we examined rectangular transmitters and receivers, each equipped with a total of $256$ antenna elements, and assessed achievable rates across various 3D rotation angles. Additionally, we varied the parameters $N_{s,\text{v}}$ and $N_{s,\text{h}}$ while maintaining a constant number of data streams, denoted as $N_s$. The unmarked line in the plot represents the performance of fully digital transmitters and receivers with a restricted number of data streams, serving as an upper bound for comparison. As depicted in Fig. \ref{fig:diff_Ns_n_shape}, our proposed analog-digital beam focusing scheme closely mirrors the optimal performance. In particular, the performance gap between the proposed analog-digital beam focusing scheme and the optimal beamforming scheme diminishes as the planar array reduces to the linear array.

\section{Conclusion} \label{conclusion}
In this paper, we addressed the challenge of point-to-point wide-aperture MIMO communication in LoS dominant high frequency systems, where the implementation of a traditional fully-digital array is impractical. We investigated the optimal antenna spacing to maximize achievable spectral efficiency in a hybrid wide-aperture MIMO system. This is achieved by capitalizing on the unique characteristics of the LoS channel structure, specifically by utilizing an approximated spherical wavefronts. To tackle the inherent complexity associated with solving the non-convex hybrid beamforming problem, we proposed a closed-form change-of-basis solution for the beam focusing aware hybrid precoder and combiner. This solution leverages the asymptotic characteristics of the channel gain matrix in the wide-aperture LoS channel. Our simulation results substantiated that the proposed configuration of antenna arrangement represents an optimal approach, demonstrating superior spectral efficiency gains compared to alternative arrangements. Furthermore, we presented numerical results highlighting that the proposed scheme achieves noteworthy spectral efficiency when the number of antennas is limited, and approaches its performance limit as the number of antennas increases.

\appendices
\section{Proof of Theorem \ref{thm:optimal_sapcing}}
\label{proof1}
Recall that the Frobenius norm of the matrix $\tilde{\mathbf{H}}$ is given by
\begin{align*}
    \lVert \tilde{\mathbf{H}} \rVert^{2}_{F}=NM.
\end{align*}
Consider planar arrays of which antenna element configuration is different from  \eqref{eqn:optnonpara} but satisfies  
\begin{align} \begin{split} \label{eqn:nonopt}
    &\cos \theta_r \sin \phi_r \bar{r}^{x}_{n} + \sin \theta_r \bar{r}^{y}_{n} - \cos \theta_r \cos \phi_r (D+\bar{r}^{z}_{n}) = 0,  \\
    &\cos \theta_t \sin \phi_t \bar{t}^{x}_{m} + \sin \theta_t \bar{t}^{y}_{m} + \cos \theta_t \cos \phi_t \bar{t}^{z}_{m} = 0,
\end{split}
\end{align}
which are the plane equations for Tx and Rx in Fig. \ref{fig:systemmodel}, respectively. With a slight abuse of the notation, let us define a matrix $\bar{\mathbf{H}}$ for the planar array in \eqref{eqn:nonopt} as $\left[\bar{\mathbf{H}}\right]_{n,m} = e^{j \frac{2 \pi}{\lambda} \frac{\bar{r}^{x}_{n} \bar{t}^{x}_{m} + \bar{r}^{y}_{n} \bar{t}^{y}_{m}}{D}}$ with $\lVert \bar{\mathbf{H}} \rVert^{2}_{F} = NM$ and $\bar{\boldsymbol{\lambda}}$ as the corresponding eigenvalues of the matrix $\bar{\mathbf{H}}$. From Lemma \ref{lma:lma1},  the singular value distribution of $\bar{\mathbf{H}}$ reveals that either the channel power is spread over more than $N_s$ singular values, i.e., $\sum^{N_s-1}_{i=0} \bar{\lambda}_{i} < NM$, or the channel power is focused on $N_s$ parallel streams while the channel power is unequally distributed over $N_s$ singular values, that is, $\bar{\lambda}_{0} \ge \bar{\lambda}_{1} \ge \ldots \bar{\lambda}_{N_s - 1}$. For such a MIMO system, the corresponding spectral efficiency is given as
\begin{align}
    \sum_{i}^{N_s} \left[ \log \left(\mu(\bar{\boldsymbol{\lambda}}) \bar{\lambda}_{i}\right) \right]^{+}.
\end{align}
Assuming that every $N_s$ sub-channel is allocated non-zero transmit power, that is, $\log \mu(\bar{\boldsymbol{\lambda}}) \bar{\lambda}_{i} \ge 0$ for all $i \in \{0, \ldots N_s-1\}$, we obtain the following inequality:
\begin{align} \label{eqn:NSproof}
\begin{split}
    N_s \sum_{i}^{N_s} \frac{1}{N_s} \log \left(\mu(\bar{\boldsymbol{\lambda}}) \bar{\lambda}_{i}\right) \le N_s \log \left( \frac{P_{\text{t}}}{N_s} \cdot \frac{\sum_{i}^{N_s} \bar{\lambda}_{i}}{N_s}\right)
\end{split},
\end{align}
where equality holds when $\bar{\lambda}_{0} = \cdots = \bar{\lambda}_{N_s -1}$. Due to the fact that the right-hand side of \eqref{eqn:NSproof} is always smaller than or equal to $N_s \log \left( \frac{P_{\text{t}}}{N_s} \frac{NM}{N_s}\right)$, the maximum achievable spectral efficiency using the antenna configuration corresponding to $\bar{\mathbf{H}}$ is less than the maximum achievable spectral efficiency of the optimal antenna configuration in \eqref{eqn:optnonpara}.

On the other hand, for the case if there exist some sub-channels that are not allocated transmit power, let us assume that for $k$ eigenvalues $\log \left( \mu(\bar{\boldsymbol{\lambda}}) \bar{\lambda} \right)$ are less than zero, then the following inequality holds.
\begin{align} \label{eqn:case2}
\begin{split}
    \sum_{i}^{N_s} \left[ \log \mu(\bar{\boldsymbol{\lambda}}) \bar{\lambda}_{i}\right]^{+} \le \left( N_s - k\right) \log \left( \frac{P_{\text{t}}}{N_s-k} \cdot \frac{N_t N_r}{N_s-k} \right)
\end{split}.
\end{align}
The right-hand side of \eqref{eqn:case2} is smaller than or equal to $N_s \log \left( \frac{P_{\text{t}}}{N_s} \cdot \frac{N_t N_r}{N_s} \right)$ if the function $x \log (\frac{P_{\text{t}}NM}{x^2})$ is increasing with respect to $x$ over the interval $N_s -k \le x \le N_s$, which typically holds for a hybrid architecture in THz MIMO systems where $N_s \ll N_tN_r$ over wide range of transmit power. In consequence, the maximum achievable spectral efficiency using the non-optimal configuration is less than the maximum achievable spectral efficiency of the proposed optimum antenna configuration. Therefore, for any case, we can conclude that an antenna configuration other than the proposed optimum configuration results in a lower achievable spectral efficiency.

\section{Array Aperture Constraint} \label{aperture_proof}
The consequence of Corollary \ref{thm:cor1} is that the spectral efficiency maximizing array configuration satisfies 
\begin{align} \label{eqn:SEmaxd}
d^{\star}_{\text{t,v}}d^{\star}_{\text{r,v}}d^{\star}_{\text{t,h}}d^{\star}_{\text{r,h}}NM =  N_s \lambda_c^2 D^2.
\end{align}
The array aperture of the rectangular array that deploys $N_{\text{v}}, N_{\text{h}}$ antennas with inter-antenna spacing $d_{\text{r,v}}, d_{\text{r,h}}$ is approximately given as $\sqrt{ \left( N_{\text{v}}d_{\text{r,v}} \right)^2 +  \left( N_{\text{h}}d_{\text{r,h}} \right)^2}$ according to the definition given in \eqref{eqn:aperture}. The aperture of the transmitter array can be derived similar to that of the receiver. To figure out the minimum aperture size that enables the deployment of \eqref{eqn:SEmaxd}, we use the arithmetic-geometric mean inequality to derive the minimum value of $\sqrt{ \left( N_{\text{v}}d_{\text{r,v}}\right)^2 +  \left( N_{\text{h}}d_{\text{r,h}}\right)^2}$ as
\begin{align}
    \sqrt{ \left( N_{\text{v}}d_{\text{r,v}}\right)^2 +  \left( N_{\text{h}}d_{\text{r,h}}\right)^2} \ge \sqrt{2 N_{\text{v}} N_{\text{h}} d_{\text{r,v}}d_{\text{r,h}}}.
\end{align}
From the above inequality, we can draw that the optimal configuration is formidable if
\begin{align*}
    L_{\text{t}} < \sqrt{2 M_{\text{v}} M_{\text{h}} d^{\star}_{\text{t,v}} d^{\star}_{\text{t,h}}} \text{ and } L_{\text{r}} < \sqrt{2 N_{\text{v}} N_{\text{h}} d^{\star}_{\text{r,v}} d^{\star}_{\text{r,h}}},
\end{align*}
for every possible $d^{\star}$s that satisfy \eqref{eqn:SEmaxd}. That is, the product of the transmitter aperture and receiver aperture should be larger than $2 \sqrt{N_s} \lambda_c D$ to configure the optimal LoS MIMO system; otherwise it is impossible.

\section{Proof of Theorem \ref{thm:Hgain}} \label{proof2}
First, we show that the channel gain matrices are a Hermitian doubly block Toeplitz matrix. The $(i,k)$ element of the channel gain matrix $\mathbf{G}_{\text{t}}$ is
\begin{align*}
    [\mathbf{G}_{\text{t}}]_{i,k} &= \sum_{l=1}^{N-1} [\tilde{\mathbf{H}}^{*}]_{i,l} [\tilde{\mathbf{H}}]_{l,k} \\
    &= \sum_{l=0}^{N-1} \left([\tilde{\mathbf{H}}]_{l,i}\right)^{*} \left([\tilde{\mathbf{H}}]_{l,k}\right)\\
    &= \sum_{l_\text{v}=0}^{N_{\text{v}}-1} \sum_{l_\text{h}=0}^{N_{\text{h}}-1} e^{-j2\pi (\alpha l_\text{v} i_\text{v} + \beta l_\text{h} i_\text{h} - \gamma l_\text{h} i_\text{v} - \eta l_\text{v} i_\text{h})} \\
    &\quad\cdot e^{j2\pi (\alpha l_\text{v} k_\text{v} + \beta l_\text{h} k_\text{h} - \gamma l_\text{h} k_\text{v} - \eta l_\text{v} k_\text{h}) } \\
    &= \sum_{l_\text{v}=0}^{N_{\text{v}}-1} \sum_{l_\text{h}=0}^{N_{\text{h}}-1} e^{-j2\pi (\alpha l_\text{v} (i_\text{v} - k_\text{v})+ \beta l_\text{h} (i_\text{h} - k_\text{h})} \\ 
    &\quad\cdot e^{j2\pi \gamma l_\text{h} (i_\text{v} - k_\text{v}) + \eta l_\text{v} (i_\text{h}-k_\text{h}))} \\
    &= \sum_{l_\text{v}=0}^{N_{\text{v}}-1} e^{-j2\pi (\alpha l_\text{v} (i_\text{v} - k_\text{v}) - \eta l_\text{v} (i_\text{h}-k_\text{h}))} \\ 
    &\quad\cdot
    \left(\sum_{l_\text{h}=0}^{N_{\text{h}}-1} e^{-j2\pi (\beta l_\text{h} (i_\text{h} - k_\text{h}) - \gamma l_\text{h} (i_\text{v} - k_\text{v}))}\right) \\
    &= \sum_{l_\text{v}=0}^{N_{\text{v}}-1} e^{-j2\pi (\alpha l_\text{v} (i_\text{v} - k_\text{v}) - \eta l_\text{v} (i_\text{h}-k_\text{h}))} \\ 
    &\quad\cdot \frac{1-e^{-j2\pi(\beta(i_\text{h} - k_\text{h}) - \gamma(i_\text{v} - k_\text{v}))N_{\text{h}}}}{1-e^{-j2\pi(\beta(i_\text{h} - k_\text{h}) - \gamma(i_\text{v} - k_\text{v}))}} \\
    &= \frac{1-e^{-j2\pi(\beta(i_\text{h} - k_\text{h}) - \gamma(i_\text{v} - k_\text{v}))N_{\text{h}}}}{1-e^{-j2\pi(\beta(i_\text{h} - k_\text{h}) - \gamma(i_\text{v} - k_\text{v}))}} \\ 
    &\quad\cdot \frac{1-e^{-j2\pi(\alpha(i_\text{v} - k_\text{v}) - \eta(i_\text{h}-k_\text{h}))N_{\text{v}}}}{1-e^{-j2\pi(\alpha(i_\text{v} - k_\text{v}) - \eta(i_\text{h}-k_\text{h}))}},    
\end{align*}  
where we have represented the antenna configuration parameter as $\alpha, \beta, \gamma, \eta$ for the notational convenience. The channel gain matrices of the uniform planar array are a Hermitian doubly block Toeplitz matrix regardless of the antenna configuration (e.g. arbitrary tilt and rotation, antenna spacing, form), with different parameters $\alpha, \beta, \gamma, \eta$. For example,
\begin{align*}
    &\alpha = \frac{d_{\text{t,v}}d_{\text{r,v}}}{\lambda_c D}, \beta = \frac{d_{\text{t,h}}d_{\text{r,h}}}{\lambda_c D}, \gamma = \frac{d_{\text{t,v}}d_{\text{r,h}}}{\lambda_c D}, \eta = \frac{d_{\text{t,h}}d_{\text{r,v}}}{\lambda_c D},
\end{align*}
are antenna configuration parameters corresponding to the optimal array investigated in Section \ref{optspacing}. Each element of the matrix $\mathbf{G}_{\text{t}}$  depends only on $(i_\text{h}-k_\text{h})$ and $(i_\text{v}-k_\text{v})$. That is, the matrix $\mathbf{G}_{\text{t}}$ consists of $M_{\text{v}} \times M_{\text{v}}$ blocks with Toeplitz structure, each block being a Toeplitz matrix of dimension $M_{\text{h}} \times M_{\text{h}}$. It is straightforward to see that $[\mathbf{G}_{\text{t}}]_{i,k} = [\mathbf{G}_{\text{t}}]^{*}_{k,i}$, thus $\mathbf{G}_{\text{t}}$ is a Hermitian doubly block Toeplitz matrix. Similar procedure proves $\mathbf{G}_{\text{r}}$ is indeed a Hermitian doubly block Toeplitz matrix.

According to \cite{gutierrez2012block}, the doubly block Toeplitz matrix is asymptotically equivalent to the corresponding doubly block circulant matrix, both generated by the same generating function. Consequently, the doubly block Toeplitz matrix can be asymptotically diagonalized by the 2D-DFT matrix. It is important to note that, while the notion of asymptotic diagonalization is widely accepted in the literature, a formal proof has not been provided to date.

%
%

\ifCLASSOPTIONcaptionsoff
  \newpage
\fi

\bibliographystyle{IEEEtran}

\bibliography{IEEEabrv,references}

\end{document}